\documentclass[prb,preprintnumbers,superscriptaddress,amsmath,amssymb,floatfix,aps,twocolumn]{revtex4}
\usepackage{graphicx}

\usepackage{lipsum} 
\usepackage{comment}
\usepackage{lineno}

\usepackage{xcolor}

\newcommand{\bluemark}[1] {\color{black}#1\color{black}\normalsize}

\usepackage{ulem}

\usepackage{afterpage}

\usepackage[version=4]{mhchem}

\usepackage{hyperref}
\usepackage{amsmath}
\usepackage{amsbsy}
\usepackage{float}
\usepackage[utf8x]{inputenc}

\begin{document}

\title{Altermagnetism in Two-Dimensional {Ca}$_2${RuO}$_4$ perovskite}

\date{\today} 

\author{J. W. Gonz\'alez}
\email{jhon.gonzalez@uatonf.cl}
\affiliation{Departamento de Física, Universidad de Antofagasta, Av. Angamos 601, Casilla 170, Antofagasta, Chile}


\author{A.M León}
\affiliation{Departamento de F\'{i}sica, Facultad de Ciencias, Universidad de Chile, Casilla 653, Santiago, Chile.}
\affiliation{Institute for Solid State and Materials Physics, TU Dresden University of Technology, 01062 Dresden, Germany}

\author{C. Gonz\'alez-Fuentes}
\affiliation{Instituto de Física, Pontificia Universidad Católica de Chile, Vicuña Mackena 4860, 7820436 Santiago, Chile}

\author{R. A. Gallardo}
\affiliation{Departamento de Física, Universidad Técnica Federico Santa María, Avenida España 1680, Valparaíso, Chile}

\begin{abstract}

We propose and characterize a novel two-dimensional material, 2D-CRO, derived from bulk calcium-based ruthenates (CROs) of the Ruddlesden-Popper family, Ca$_{n+1}$Ru$_n$O$_{3n+1}$ ($n = 1$ and $2$). Using density functional theory, we demonstrate that 2D-CRO maintains structural stability down to the monolayer limit, exhibiting a tight interplay between structural and electronic properties. Notably, 2D-CRO displays altermagnetic behavior, characterized by zero net magnetization and strong spin-dependent phenomena, stabilized through dimensionality reduction. This stability is achieved by breaking inversion symmetry along the z-axis, favoring altermagnetic properties even in the absence of van der Waals interactions. 
Through theoretical models and computational analysis, we explore the altermagnetic behavior of 2D-CRO, both with and without spin-orbit coupling. We identify the spin components that contribute to the altermagnetic character and highlight the potential of 2D-CRO as a promising material for investigating altermagnetic phenomena and topological features.

\end{abstract}

\maketitle
 
\section{\label{sec:intro} Introduction}

The recent interest in two-dimensional (2D) crystals has shifted towards antiferromagnetic (AF) materials. Unlike ferromagnetic (FM) materials, AF materials have zero net magnetization and are robust against perturbations from magnetic fields. They are expected to enable information writing three orders of magnitude faster than ferromagnetic-based systems, providing a promising platform for spin control in the femtosecond regime  \cite{baltz2018antiferromagnetic,rahman2021recent,vsmejkal2018topological}. 
Unfortunately, the absence of strong spin-dependent effects has limited their practical applicability. Recently discovered altermagnetic (AM) materials offer zero net magnetization combined with the strong spin-dependent phenomena characteristic of ferromagnets, opening new possibilities for overcoming the challenges faced by conventional AF spintronic materials \cite{vsmejkal2022emerging,vsmejkal2018topological,krempasky2024altermagnetic}. 
\bluemark{Earlier studies on the role of microscopic multipole moments have provided an alternative theoretical foundation for understanding altermagnetic phenomena. Their findings highlight the role of antisymmetric spin splitting driven by non-relativistic interactions, which aligns conceptually with the modern understanding of altermagnetic behavior \cite{hayami2019momentum,hayami2020bottom,hayami2020spontaneous}}.
The altermagnetic field has paved the way for a new branch in the search for magnetic materials, 
encompassing 3D and 2D systems of different natures. To date, a large family 
of correlated materials has been proposed to display 
AM features \cite{osumi2024observation,fedchenko2024observation}, 
most of them in bulk form. Only recently has the exploration of AM features expanded 
to include 2D materials \cite{mazin2023induced,liu2024twisted}.

Since altermagnetism strongly depends on magnetic space groups, it exhibits non-straightforward behavior on surfaces where dimensionality reduction and symmetry changes occur \cite{sattigeri2023altermagnetic}. Unlike common antiferromagnetic 2D materials, altermagnetism may not be viable in van der Waals (vdW) systems \cite{mazin2023induced}. The primary challenge identified in 2D systems for achieving AM states is their inherent high symmetry, which necessitates mechanisms to induce inversion symmetry breaking. In this context, exploring beyond vdW materials, such as non-vdW materials, can expand the search for AM states into 2D systems. Unlike vdW systems, non-vdW materials with uncompensated bonds or in-plane distortions can easily break in-plane symmetries, favoring the emergence of AM features. However, these aspects still need to be explored.

This work proposes a novel 2D material with the stoichiometry \ce{Ca2RuO4}, derived from calcium-based ruthenates (CROs) of the Ruddlesden-Popper family, Ca$_{n+1}$Ru$_n$O$_{3n+1}$ ($n = 1$ and $2$). These systems are well-known for exhibiting diverse phenomena, including colossal magnetoresistance \cite{ohmichi2004colossal}, spin waves \cite{kikugawa2010ca3ru2o7}, and superconductivity under pressure \cite{alireza2010evidence}, among others \cite{dietl2018tailoring}. In their ground state, Ca$_2$RuO$_4$ exhibits a Mott-insulating antiferromagnetic (AF) state \cite{dietl2018tailoring}, while Ca$_3$Ru$_2$O$_7$ is an AF polar metal \cite{markovic2020electronically}. Their distinct electronic properties arise from slight structural differences along the c-axis, leading to different rotational degrees of freedom of RuO$_6$ octahedra. Despite the strong influence of layer dimensionality on the electronic and magnetic properties of these systems, they are cataloged as quasi-2D systems, as their in-plane behavior dictates their exotic electronic properties \cite{markovic2020electronically,bertinshaw2019unique}. In this context, exploring their monolayer forms may reveal the effects of thickness reduction on electronic characteristics and provide novel perspectives for discovering or tuning their electronic phases.

Here, we show that Ca$_{n+1}$Ru$_n$O$_{3n+1}$  exhibits structural stability down to the 2D limit (2D-CRO), like its bulk counterpart \ce{Ca3Ru2O7}
possesses correlated electronic and structural properties.
Furthermore, 2D-CRO exhibits an altermagnetic magnetic character similar to what was reported for the \ce{Ca2RuO4} bulk system \cite{cuono2023orbital}; however, the altermagnetic features are enhanced through dimensionality reduction. To further understand the role of symmetry and symmetry operations, we propose a toy model based on the minimal components and test the role of the inversion symmetry. Our model suggests that the altermagnetic character of 2D-CRO mainly depends on the in-plane symmetry operations independent of the layer inversion symmetry.

\begin{figure*}[!]
\centering
\includegraphics[clip,width=0.9\textwidth,angle=0]{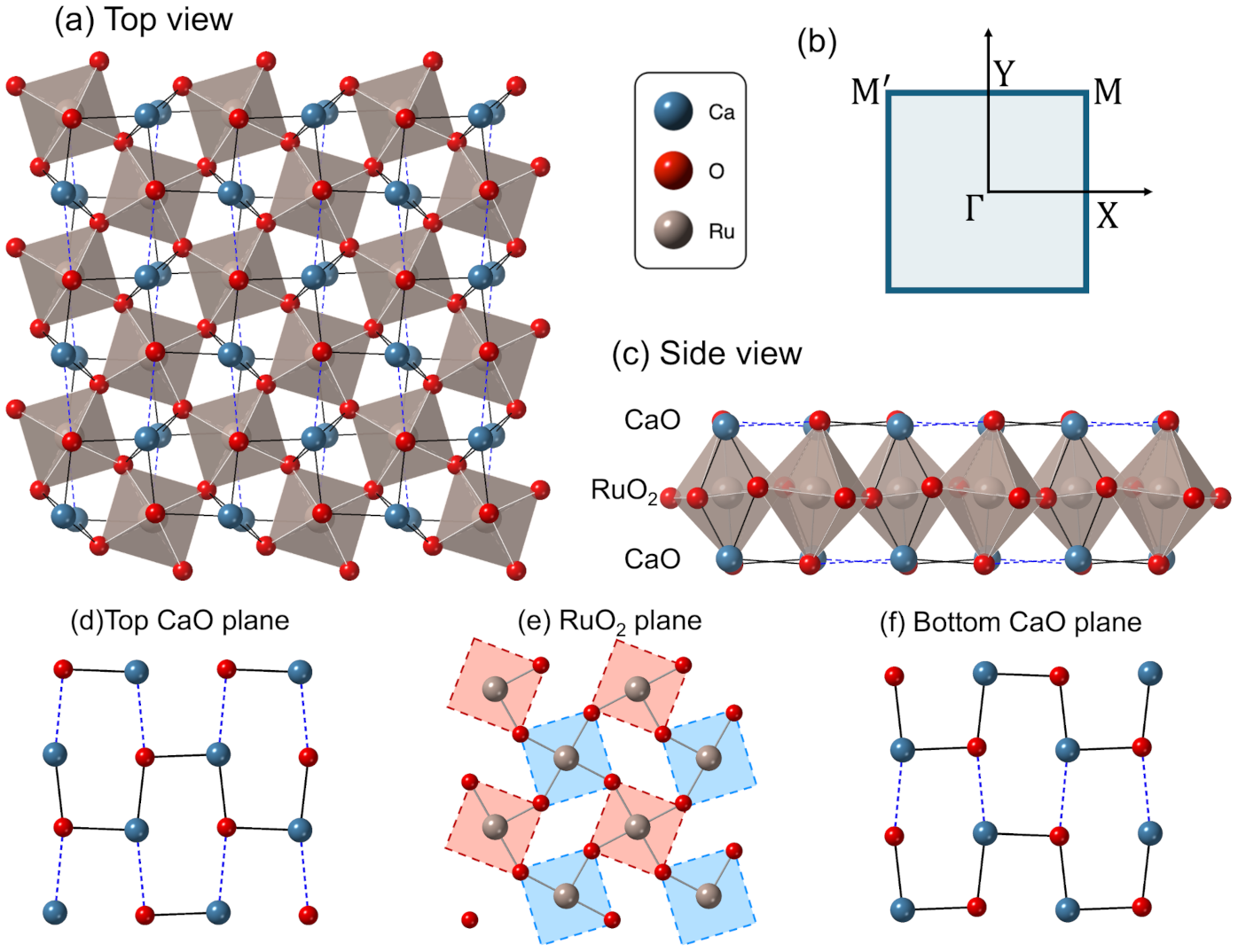} 
\caption{(Color online) Representation of \ce{Ca3Ru2O7} monolayer (2D-CRO) in the antiferromagnetic solution for U = 2 eV. 
In (a), the monolayer top view shows the Oxygen octahedral environment around Ru atoms. 
In (b), the high-symmetry points in the reciprocal space.
In (c), monolayer side view. Note the presence of two types of Ca-O, one shorter (as black lines) and one larger (as blue-dashed lines).
In (d-f), we present a top view of each atomic plane from 2D-CRO in a $2\times 2$ supercell.
In panels (d) and (f), we show the alternating CaO bonds forming an armchair pattern.
In panel (e), we show the \ce{RuO2} plane; as an eye guide, we include blue and red squares to identify the two types of Ru environments.
}
\label{Fig:scheme}
\end{figure*}

\section{Proposed Methods for 2D-CRO Synthesis}

\bluemark{We propose a fabrication method for synthesizing a 2D monolayer of \ce{Ca2RuO4} inspired by existing techniques for similar 2D oxide materials. Our approach combines chemical exfoliation, delamination of layered precursors, and solution-based assembly, drawing on well-established methods used for the synthesis of 2D oxide nanosheets, such as \ce{RuO2}. We expect that the proposed multi-step process may allow the production of high-quality monolayers with controlled thickness, lateral size, and crystallinity.}

\bluemark{The initial step is to select a suitable precursor for exfoliation. Based on previous work on exfoliating ruthenium-based oxides, \ce{Ca2RuO4} is a prime candidate. The bulk \ce{Ca2RuO4} is a layered perovskite structure, which allows the delamination, as demonstrated for \ce{RuO2} nanosheets \cite{ko2018understanding}. The \ce{Ca2RuO4} precursor can be synthesized as a bulk crystal using solid-state reaction methods under controlled oxygen pressure\cite{nobukane2020coappearance}.}

\bluemark{To achieve the \ce{Ca2RuO4} monolayer (2D-CRO), we propose two complementary methods: soft-chemical exfoliation and ultrasonic-assisted liquid exfoliation.
The Chemical exfoliation employs ion-exchange methods where interlayer cations in bulk \ce{Ca2RuO4} are replaced with large organic cations, such as tetrabutylammonium (TBA) or tetramethylammonium (TMA) ions \cite{sakai2024highly}. We expect the ion-exchange process to introduce water into the interlayer space, causing the structure to swell and weaken interlayer interactions. The swollen material is then subjected to agitation or shear forces to separate individual monolayer sheets\cite{ko2018understanding}.
In the ultrasonic-assisted liquid Exfoliation method, high-frequency ultrasonic waves generate cavitation bubbles in a solvent, creating intense shock waves that separate the layers of \ce{Ca2RuO4}. The process typically uses solvents like ethanol, N-methyl-2-pyrrolidone (NMP), or isopropanol\cite{wang2024recent}. The sonication process parameters, such as frequency, time, and power, are optimized to prevent the destruction of the monolayers. This approach has been widely used to exfoliate\ce{RuO2} nanosheets from K-intercalated \ce{RuO2} \cite{ko2018understanding} and to prepare MOF nanosheets \cite{wang2024recent}.}

\bluemark{To fabricate a continuous film of 2D \ce{Ca2RuO4}, we propose two techniques: Langmuir-Blodgett assembly and electrostatic self-assembly.
In the Langmuir-Blodgett a ssembly, the \ce{Ca2RuO4} nanosheets are trapped at an air-liquid interface, compressed laterally to form a dense monolayer, and transferred onto a solid substrate. This method has been applied to other oxide nanosheets \cite{sakai2024highly}.
The electrostatic self-assembly method relies on surface charge interactions. A negatively charged \ce{Ca2RuO4} nanosheet can be adsorbed onto a positively charged substrate by immersing the substrate in the colloidal suspension. This method is advantageous for creating a uniform, large-area monolayers on flexible substrates \cite{sakai2024highly}.}

\bluemark{Post-assembly annealing could improve crystallinity and remove residual solvents or organic cations. Annealing at high temperatures ($~500\, ~^\circ$C) under an inert atmosphere improves the crystallinity of \ce{RuO2} nanosheets \cite{ko2018understanding}.}

\bluemark{This proposed fabrication route for a 2D monolayer of \ce{Ca2RuO4} combines the most effective synthesis methods used for \ce{RuO2}, \ce{Ca2RuO4}, and oxide nanosheets. By employing a combination of exfoliation, delamination, and solution-based assembly, it is possible to obtain large-area, high-quality monolayers of \ce{Ca2RuO4}. }


\section{\label{sec:comp} Computational details}

We used density functional theory (DFT) with the Perdew-Burke-Ernzerhof generalized gradient approximation (PBE-GGA) functional\cite{Perdew2008}, implemented in the Vienna ab initio Simulation Package (VASP)\cite{VASP0}, for our theoretical analysis using the plane-wave pseudopotential method.

After convergence test, we set the plane-wave energy cutoff of 450 eV and use a $\Gamma$-centered Monkhorst-Pack grid of 10$\times$10$\times$1 to perform the atomic relaxations (equivalent to a $0.02\times 2\pi/$\AA{} BZ sampling) and  20$\times$19$\times$1 (equivalent to a $0.01\times 2\pi/$\AA{} BZ sampling) to perform the self-consistent calculations. 
Our structural optimization of the unit cell until a force convergence threshold of at least 10$^{-3}$ eV/\AA\ per atom.
The configuration of the valence electrons for the pseudopotentials is based on the VASP recommendations\cite{pseudos}:  O (6 valence e$^-$, including s-electrons), Ca (10 valence e$^-$), and Ru (14 valence e$^-$, including p-electrons).

Molecular dynamics simulations were conducted in a $2 \times 2 \times 1$ supercell at a constant temperature, using a reduced $\Gamma$-point scheme within an integration step of 1 fs for 10 ps. Ions are in contact to  a Nos\'e-Hoover thermostat, where the Nos\'e mass was determined by estimating the natural frequency of the system\cite{nose}.

The phonon dispersion were calculated within the harmonic approximation using
PHONOPY code\cite{togo2023first,togo2023implementation}. An accurate representation of the phonons in this approach requires a larger supercell; therefore, we chose a $2\times2\times1$ supercell.; therefore, we chose a $2\times2\times1$ supercell.

\section{Geometrical and Electronic properties}

The 2D-CRO structure can be derived from either \ce{Ca3Ru2O7} or \ce{Ca2RuO4} bulk materials. In their bulk form, these systems consist of stacked layers of \ce{RuO6} octahedra and \ce{CaO8} polyhedra, which form a honeycomb lattice along the ab-plane\cite{yoshida2005crystal,braden1998crystal}. These layers are held together by weak van der Waals forces, making them easy to exfoliate into thin layers.


Figure \ref{Fig:scheme} presents a schematic view of the 2D-CRO monolayer with the stoichiometry \ce{Ca2RuO4}. The 2D-CRO monolayer comprises two outer planes of Ca-O and an inner plane of Ru-O. The Ca-O plane features two distinct types of in-plane Ca-O bonds resulting from tilting Oxygen octahedra around the Ru atoms. In the AF configuration, these bonds measure 2.37 \AA{} and 2.77 \AA{}, while in the FM configuration, they measure 2.47 \AA{} and 2.64 \AA{}. Both electronic correlation effects and the magnetism of the system can be associated to changes in the in-plane distance between Ca and O atoms. Further details regarding these effects will be discussed in the forthcoming sections.

The central plane, composed of Ru and O atoms, can be understood as two overlapping sublattices, one sublattice associated with the environment with counterclockwise rotation (red squares in Fig. \ref{Fig:scheme}(e)) and the other sublattice with clockwise rotation (blue squares in Fig. \ref{Fig:scheme}(e)). Note that each separate sublattice is invariant under a $C_{4z}$ symmetry operation, corresponding to 90$^\circ$ rotations along the z-axis (normal to the monolayer surface). Later, we will associate these different sublattices with positive/negative projections of the antiferromagnetic coupled Ru atoms.

\begin{figure*}[!]
\centering
\includegraphics[clip,width=0.75\textwidth,angle=0]{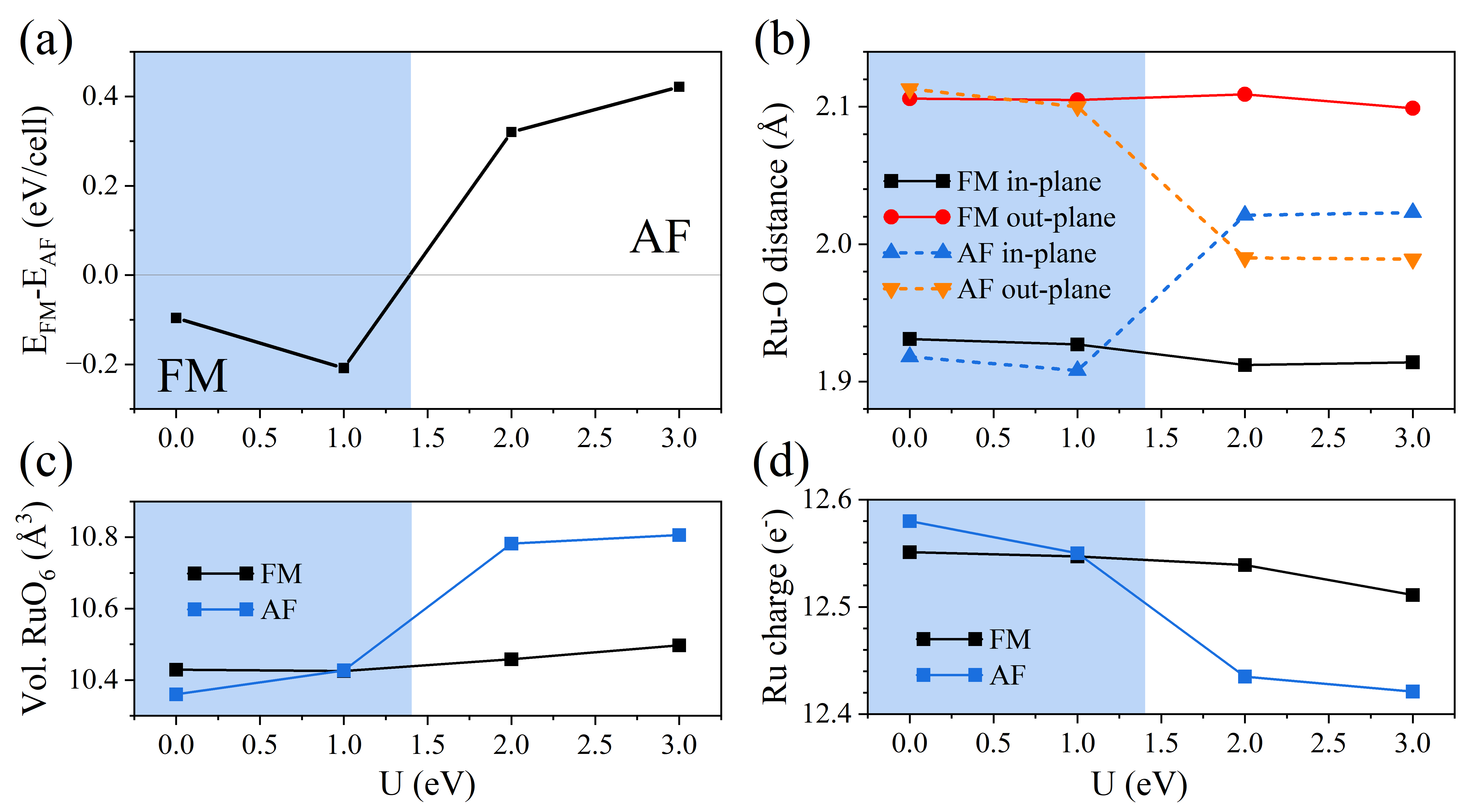} 
\caption{(Color online) Physical properties as a function of the Hubbard-U term.
In (a), the energy difference per unit cell between the antiferromagnetic (AF) 
and ferromagnetic (FM) configurations are depicted, identifying a critical 
Hubbard-U value ($U_C$) for the transition between FM and AF configurations, 
estimated to be approximately 1.4 eV. 
In (b), the Ru-O distance considers the Ru atom's first in-plane and 
out-of-plane Oxygen neighbors. In (c), the volume of the \ce{RuO6} octahedra. 
Panel (d) shows the charge per Ru atom.
}
\label{Fig:FM-AF}
\end{figure*}

The geometries used and analyzed in this study are available in the Zenodo repository\cite{Geo}.

\subsection{Electron correlation effects}

\bluemark{The effects of electronic correlation in bulk \ce{Ca2RuO4} and \ce{Ca3Ru2O7} have been extensively studied \cite{leon20243,puggioni2020cooperative,von2020resonant,sutter2017hallmarks}. These investigations reveal that the Hubbard-$U$ values applied to the $d$-orbitals of Ru atoms range from 0 to 3 eV and highlight how changes to the Hubbard-$U$ parameter can impact the magnetic behavior, electronic properties, and structural phases of the system.}

We apply the Hubbard-U correction \bluemark{to the d-orbitals of Ru atoms} in the rotationally invariant approximation introduced by Dudarev. For values between 0 and 3 eV, considering non-magnetic (NM), ferromagnetic (FM), and antiferromagnetic (AF) configurations, we completely relax the atomic positions and lattice vectors. We discard the non-magnetic solutions, which are above the ground state, +0.12 eV/cell for U = 0 eV and +1.60 eV/cell for U = 3 eV.

In Figure \ref{Fig:FM-AF}(a), we compare the energy of the FM and AF solutions as a function of the Hubbard-U term \bluemark{applied to Ru d-orbitals}. For small values of U$<U_{C}$ ($U_{C} \sim 1.4$ eV), the system presents a ferromagnetic character; for U values larger than this critical value, the system presents an antiferromagnetic character.
\bluemark{This critical Hubbard-$U$ value is comparable to that found for bulk \ce{Ca3Ru2O7}\cite{leon20243}. For U$< 1.4$ eV, the bulk \ce{Ca3Ru2O7} system presents a Bb2$_1$m symmetry where all Ru-O octahedra are similar, and for U$>1.4$ eV, the bulk \ce{Ca3Ru2O7} structural phase changes to Pn2$_1$a, presenting two types of crystallographic environments around the Ru atom. }

For the FM solutions with $U = 1 \, \text{eV}$ (GS), $U = 2 \, \text{eV}$, and $U = 3 \, \text{eV}$, the relaxed structure exhibits a triclinic crystal symmetry with space group 1 (\textit{P1}), characterized by a single symmetry operation. Similarly, the AF solution at $U = 1 \, \text{eV}$ also retains a triclinic symmetry. However, for higher $U$-values ($U = 2 \, \text{eV}$ (GS) and $U = 3 \, \text{eV}$ (GS)), the AF solutions exhibit a monoclinic crystal symmetry with space group 7 (\textit{Pc}), which includes two symmetry operations.

The crystallographic environment around ruthenium atoms is critical to understanding magnetism in CRO\cite{leon20243}. In Figure \ref{Fig:FM-AF}(b), we have the Ru-O distance to the first neighbors; we see that the distance to the in-plane and out-plane Oxygen atoms is highly dependent on the Hubbard-U parameter and the magnetic configuration of the system. For the FM configuration and U$<U_{C}$, the in-plane Ru-O distance stays around 1.93 \AA{}, while for U$>U_{C}$, the distance drops to 1.91 \AA{}. However, the Ru-O out-plane distance remains constant at 2.1 \AA{}, variations of 0.007 \AA{} between the U$ = 0$ and U$ = 3$ eV.
The antiferromagnetic systems present much more geometry variation than the FM case. For values of U$<U_{C}$, the Ru-O distances do not deviate much from the FM case. For U$>U_{C}$, there is a radical change in the geometry; now, the Ru-O in-plane distance is slightly larger, being around 0.031 \AA{} above the out-plane distance. 
This subtle change in geometry is sufficient to alter the charges and magnetic moments induced in the Oxygen atoms, later modifying the stability order of the system.

In Figure \ref{Fig:FM-AF} (d), the charge per ruthenium atom shows a behavior complementary to that shown by the \ce{RuO6} octahedron volume for values of U$<U_C$ in Fig. \ref{Fig:FM-AF}(c); for both magnetic solutions, the charge stays around $\sim 12.56$ $e^-$, where the charge difference is distributed to the in-plane Oxygen atoms. For U$>U_C$, the Ru charge in AF configuration drops to $\sim 12.42$ $e^-$, where the charge difference moves to the out-of-plane Oxygen atoms. 

\begin{figure}[!]
\centering
\includegraphics[clip,width=0.98\columnwidth,angle=0]{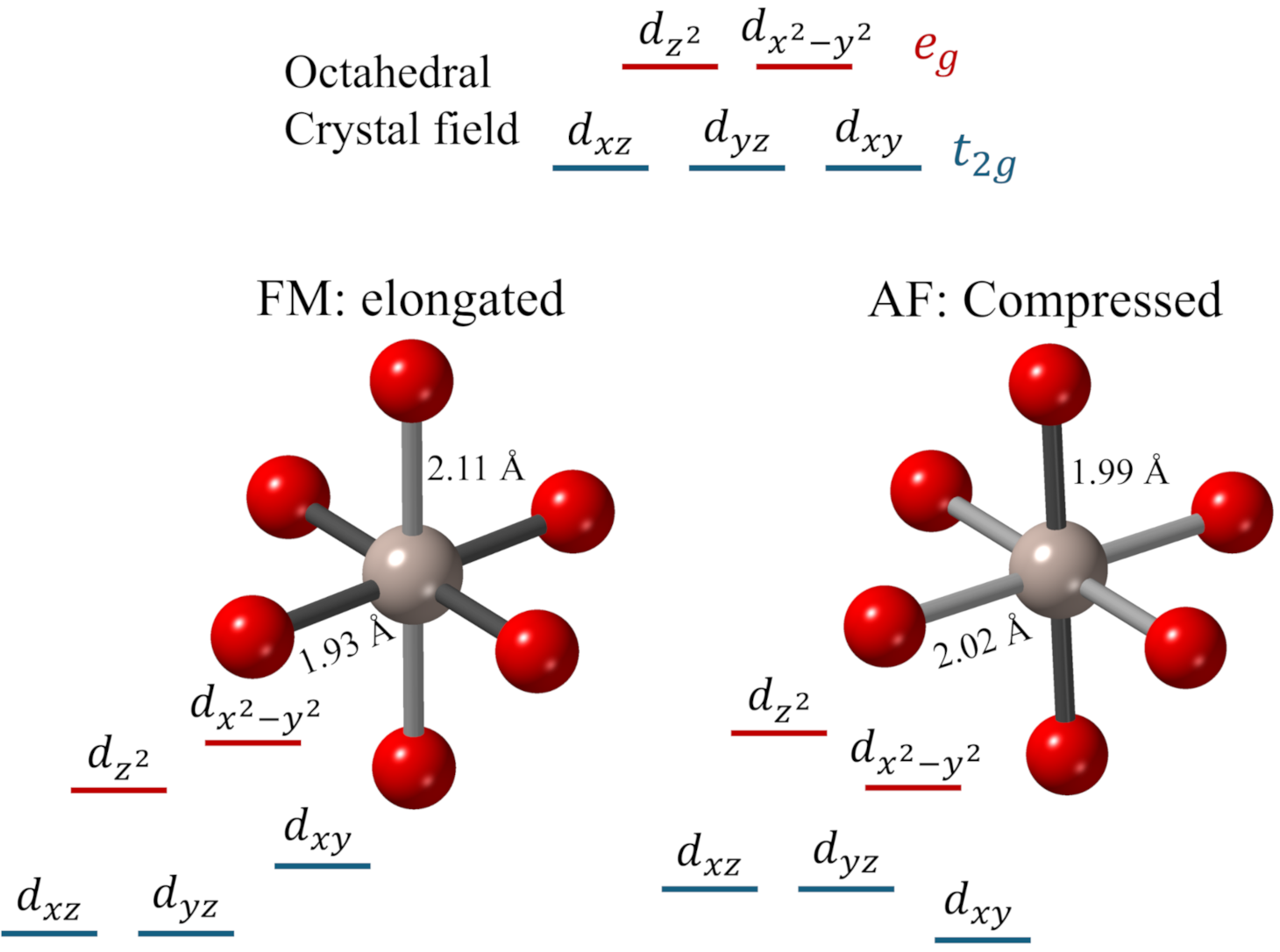} 
\caption{(Color online)  The crystallographic environment is given by the ruthenium atoms and their first six Oxygen neighbors, which form an octahedron. The top plot depicts the crystal field of a pristine octahedral environment with the $e_g$ and $t_{2g}$ d-orbitals. On the left, we have the structure of the FM system (U = 1 eV) with an elongation of the bonds in the out-of-plane direction. On the right, we have the structure of the AF system (U = 2 eV), where there is a slight compression of the out-plane distances compared to the in-plane distances.
}
\label{Fig:octa}
\end{figure}

The volume of the octahedra formed by six Oxygen atoms around the ruthenium atom is slightly modified from bulk to 2D layers. The primary effect of reducing dimensionality is slightly elongation along the z-axis, a signature of the weak layer coupling in the CRO structure. Moreover, the volume is also affected by the magnetic configuration and the Hubbard-U term, as shown in Figure \ref{Fig:FM-AF}(c) and \ref{Fig:octa}. For the FM configurations, corresponding to an elongated octahedron in the out-plane direction (details in Fig. \ref{Fig:octa}), the volume of the octahedron shows slight variation, ranging from 10.43 \AA$^3$ for U = 0 to 10.55 \AA$^3$. In the AF configurations, corresponding to a slightly compressed octahedron in the out-plane direction (Fig. \ref{Fig:octa}), we identify a regime for U$<U_C$ where the volume is 10.36 \AA$^3$, close to that shown by the FM solutions, and another regime, for U$>U_C$, where the volume increases to 10.81 \AA$^3$. 
This volume variation could be associated with the phase changes observed in the bulk of CRO\cite{leon20243}. It may also be related to the electronic redistribution in the d-shell, as described below.

The crystallographic environment is defined by an octahedron formed by the ruthenium atoms and their first six Oxygen neighbors. The top scheme in Figure \ref{Fig:octa} shows the crystal field of a perfect octahedral environment, showing the separation between orbitals with $e_g$ and $t_{2g}$ symmetries. On the left side of Fig. \ref{Fig:octa}, we have the structure of the FM system (U = 1 eV), where there is an elongation of the bonds in the out-of-plane direction, lowering the energy of the d orbitals with the z component ($d_{z^2}$, $d_{xz}$, and $d_{yz}$). On the right side of Fig. \ref{Fig:octa}, we have the structure of the AF system (U = 2 eV), where there is a slight compression of the out-plane distances compared to the in-plane distances; this slight compression causes the $d_{x^2-y^2}$ and $d_{xy}$ orbitals to shift down in energy slightly.  

Table \ref{Tab:a} summarizes lattice constants, lattice constant ratios (a/b), and the magnetic moments of ruthenium and Oxygen atoms. The magnetic moments of the ruthenium and in-plane Oxygen atoms are slightly higher in the ferromagnetic solutions. In the antiferromagnetic configurations, the out-of-plane Oxygen atoms have a higher magnetic moment than their in-plane counterparts, likely due to the enhancement of the \ce{RuO6} volume octahedra, where the Ru-O$_{out}$ bonds become shorter (see Fig. \ref{Fig:FM-AF}(d)). 

\bluemark{The Bader charge analysis for Ruthenium atoms (Figure \ref{Fig:FM-AF}(d)) and Oxygen atoms (Figure \ref{Fig:bader}) for the FM configurations reveals minimal variation in the charge of Ru and O atoms, both in-plane and out-of-plane atoms. This behavior is consistent with the small changes in Ru-O bond lengths across different values of \( U \), as shown in Fig.~\ref{Fig:FM-AF}(b). 
In the AF case, for \( U \) values below the critical value (\( U_C \sim 1.4 \, \text{eV} \)), the atomic charges and Ru-O bond lengths are comparable with those of the FM configuration. 
However, differences appear for \( U > U_C \), where the AF solution becomes the ground state. Specifically, we observe a reduction in the charge of the Ru and in-plane O atoms (Fig.  \ref{Fig:bader}) , accompanied by a decrease in the Ru-O in-plane bond length (Figure \ref{Fig:FM-AF}(d)). Conversely, the out-of-plane O charge increases together with the Ru-O out-of-plane bond length. 
These findings correlate strongly with the bond-length variations induced by the Hubbard-U parameter and magnetic configuration (FM vs. AF), as shown in Fig.~\ref{Fig:FM-AF}(b).
Our analysis suggests that the stabilization of the AF solutions arises from a combination of changes in atomic charges, variations in Ru-O bond lengths, and the resulting rotations and tilting of the octahedra.}

\subsection{Magnetic exchange coupling}
To analyze the interactions between magnetic atoms, we used second-order perturbation 
theory combined with the magnetic force theorem (MFT), as implemented in the OpenMX 
code\cite{yoon2018reliability,ozaki2003variationally}. 
For OpenMX calculations, we use parameters that are equivalent to those with VASP. 
Later, we use the JX\cite{yoon2020jx,han2004electronic} 
and TB2J\cite{he2021tb2j} packages to post-process the OpenMX output and calculate 
the magnetic exchange coupling parameters. The results obtained from both packages 
were consistent, indicating reliable exchange values. 
Note that this approximation has been demonstrated to be reliable, yielding values 
close to experimental data \cite{leon2020strain,han2004electronic,pizzochero2020magnetic}.

Using the AF solution for a Hubbard-U = 2 eV applied to Ru atoms, we determine an effective ruthenium-ruthenium first-neighbor magnetic coupling of $J_{Ru-Ru} = 3.10$ meV.
We prepare an Ising model with these parameters, considering only magnetic interactions between ruthenium atoms at first neighbors. Assuming an FM configuration with the value, we find a Curie temperature of 85.45 K, which is in the same order as the experimental report for \ce{Ca3Ru2O7} bulk and \ce{Ca2RuO4} nanofilms \cite{yoshida2005,nobukane2020co}. 
Using the Ising model with first-neighbor interactions in an AF scenario, we estimate a Néel temperature of 121.4 K. This value is in good agreement with the experimental result of $110 \pm 10$ K reported for antiferromagnetic \ce{CaRuO3} crystals \cite{longo1968magnetic}.

Further details in section \ref{sec:ising}.

\subsection{Altermagnetic properties}

\begin{figure}[t!]
\centering
\includegraphics[clip,width=1\columnwidth,angle=0]{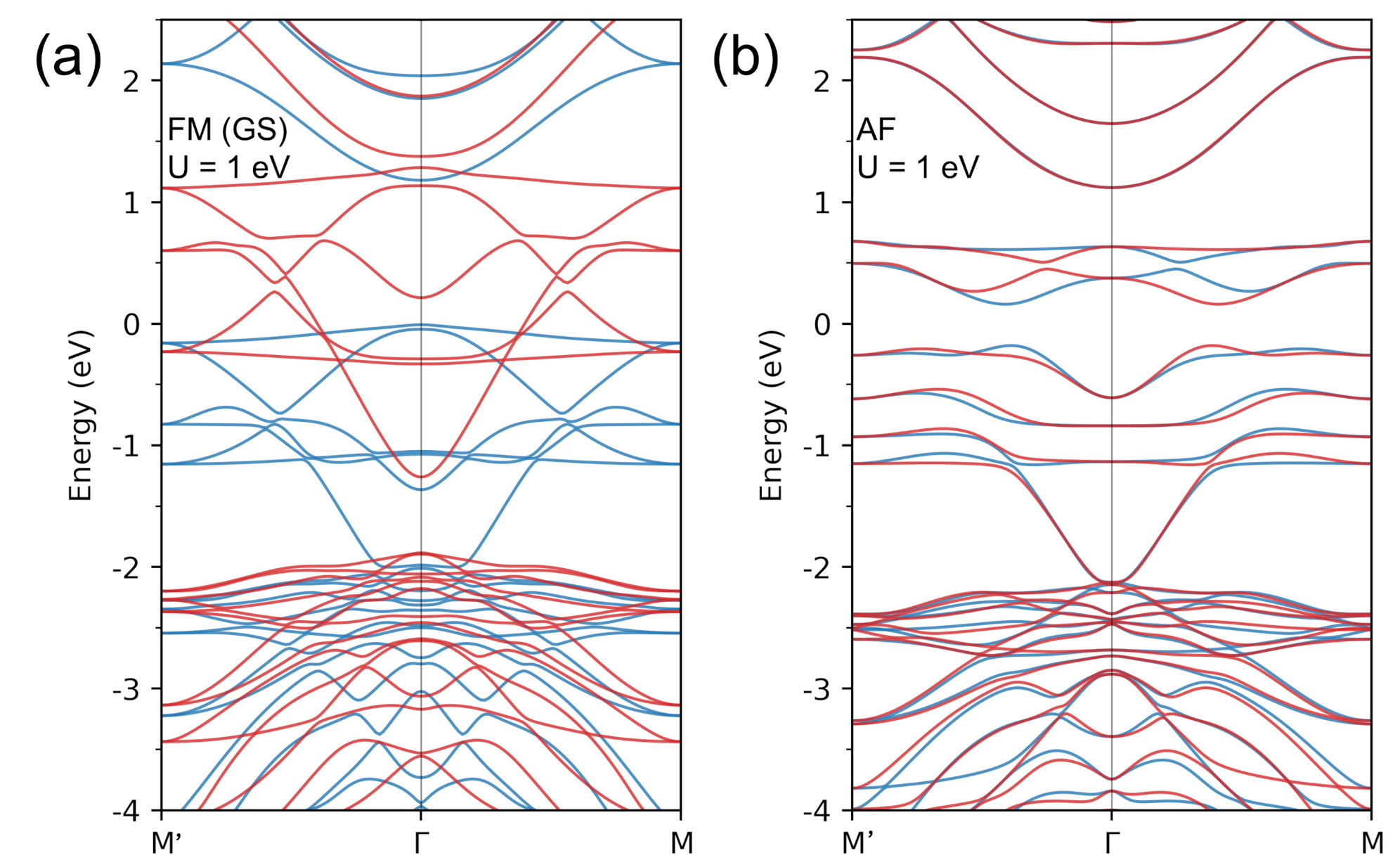} 
\caption{(Color online) Band structure with U = 1 eV. Panel (a) shows the ferromagnetic (FM) configuration, while panel (b) displays the antiferromagnetic (AF) configuration. The blue and red lines indicate the spin-up and spin-down components, respectively. Note that the band structure in the FM system (panel a) is symmetric with respect to the $\Gamma$-point, whereas this symmetry is broken in the AF system (panel b).
}
\label{Fig:bandsU1}
\end{figure}

The band structure in Fig. \ref{Fig:bandsU1} reveals the ground state at  U$ = 1$ eV as a semimetal with a ferromagnetic (FM) coupling between the Ru atoms, followed by the antiferromagnetic (AF) configuration, which presents a narrow-gap semiconductor character.
At U = 1 eV (Fig. \ref{Fig:bandsU1} (b)), we note that the AF band structure displays k-dependent spin-splitting, a characteristic feature of altermagnetic (AM) materials\cite{vsmejkal2022emerging,sattigeri2023altermagnetic,jungwirth2016antiferromagnetic}. 
This altermagnetic character is evident from comparing the up/down bands near the Fermi level along the M'-$\Gamma$ and $\Gamma$-M paths.

\begin{figure*}[t!]
\centering
\includegraphics[clip,width=.95\textwidth,angle=0]{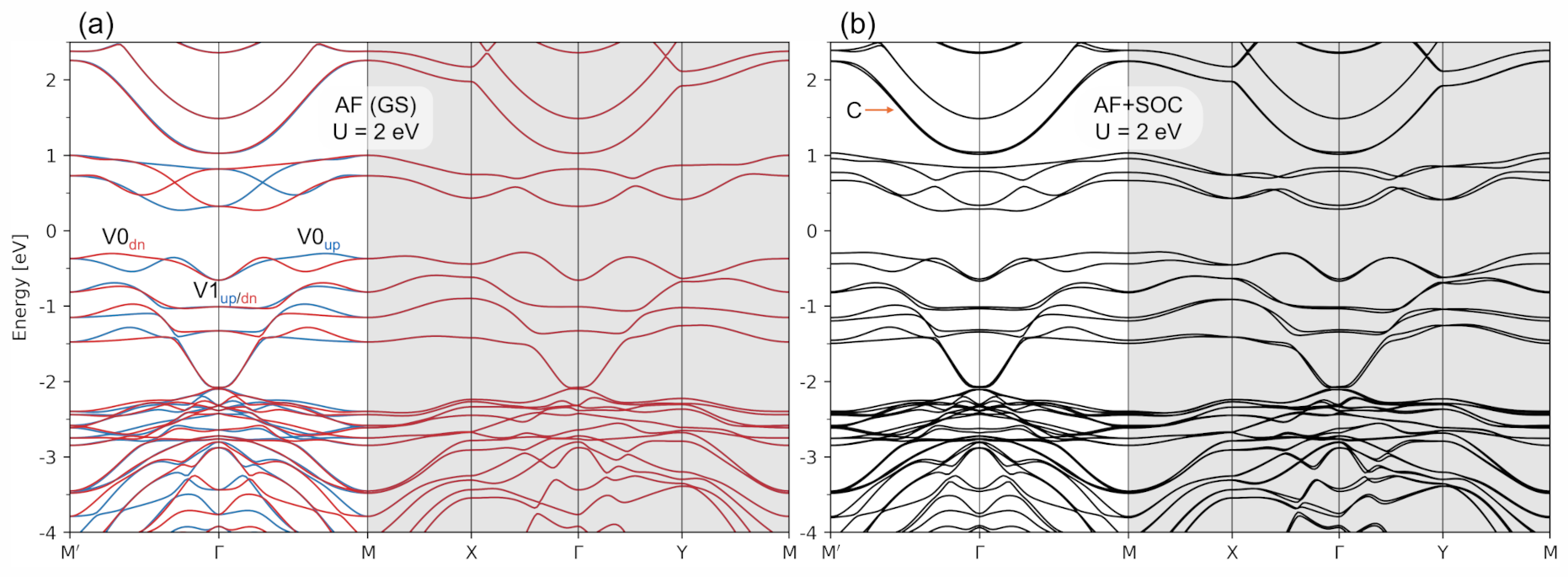} 
\caption{(Color online) The band structure for the antiferromagnetic 
system with U = 2 eV. In (a), blue and red lines indicate the spin up 
and down components; in (b), the bands with spin-orbit coupling.
\bluemark{The valence bands V0 corresponds to the highest valence band and V1 to the second-highest valence band. The band labeled as C is a doubly degenerate conduction $e_g$-band, which we follow as a function of the applied electric field, details in Fig. \ref{Fig:electric_bands}. Note that this band structure corresponds to an extended path of Fig.~\ref{Fig:bandsU1}(a) where the system only presents altermagnetic character along the M'$-\Gamma-$M path.} 
}
\label{Fig:bandsU2}
\end{figure*}

When we increase the value of the Hubbard-U term, where the most stable configuration corresponds to the solution with AF coupling between the ruthenium atoms, in the AF band structure for U = 2 eV (Fig. \ref{Fig:bandsU2}(a)), the k-dependent band splitting along M'-$\Gamma$-M points becomes more evident, reaching values close to $\pm 0.2$ eV for the highest occupied valence band.

Note that the band structure in Figures \ref{Fig:bandsU1} (b) and \ref{Fig:bandsU2}(a) is not the trivial 
AF behavior where, according to the Kramers spin degeneracy 
theorem \cite{kramers1930theorie,vsmejkal2018topological}, AF bands are expected to have degeneracy 2.

\bluemark{To understand the orbital contributions to the V0 and V1 bands (Figure \ref{Fig:bandsU2}(a)), we performed a detailed analysis of the orbital-projected band structure, as shown in Figure \ref{Fig:pband_AF}. Our results reveal that the V0 and V1 bands originate from a hybridization of $t_{2g}$ orbitals ($d_{xy}$, $d_{xz}$, $d_{yz}$). 
To properly disentangle the contributions of these orbitals, we rotated the crystal structure by 45$^\circ$ around the $z$-axis, aligning the internal octahedral crystal field with the Cartesian axes\cite{hu2017chemical,rassekh2020remarkably}. This rotation provides a clearer decomposition of the orbital components contributing to the band structure. The observed orbital hybridization is consistent with the crystal field effects of Ru-O octahedra, a common feature in transition-metal oxides (see Fig. \ref{Fig:octa}).}

To understand the $k$-dependent band splitting along the M'-$\Gamma$-M path, it is essential to recall the geometric arrangement of 2D-CRO. The plane containing the Ru and O atoms can be described as two overlapping sublattices, each with opposite magnetic moments. 
One of these sublattices is characterized by a counterclockwise rotation of the Ru-O octahedra (depicted by red squares in Fig. \ref{Fig:scheme}(e)), while the other sublattice exhibits a clockwise rotation (represented by blue squares in Fig. \ref{Fig:scheme}(e)). Despite these opposite rotations, each sublattice remains invariant under the $C_{4z}$ symmetry operation, corresponding to a 90$^\circ$ rotation around the $z$-axis (normal to the monolayer surface).

To understand the role of symmetry in 2D altermagnets, we consider the effect of a time-reversal operation, which exchanges spin-up and spin-down states. Unlike traditional antiferromagnets, in 2D-CRO, it is not possible to recover the original magnetic configuration using a simple translation. 
Instead, to restore the original structure after a time-reversal operation, a $2_1$ screw symmetry operation along the $x$-axis must be applied. This operation combines a $C_{2x}$ rotation (equivalent to $y \rightarrow -y$ and $z \rightarrow -z$ inversions) with a half-unit cell translation along the $x$-axis, resulting in $x \rightarrow x + 0.5$. The combined action of rotation and translation introduces $k$-dependent splitting of electronic states, as described in recent studies \cite{vsmejkal2022emerging}. 
We performed the symmetry analysis using the \texttt{pymatgen} library \cite{ong2013python}, which allowed us to identify the key symmetry elements of the system. Additional details regarding the role of symmetry in the electronic properties can be found in Section \ref{ToyModel}.

\section{Spin-Orbit effects}
We first determine the magnetic anisotropy energy (MAE), which is calculated as the energy difference between magnetic configurations with different orientations of the magnetic moments. Figure \ref{Fig:MAE} presents the MAE as a function of the azimuthal ($\phi$) and polar ($\theta$) angles, clearly showing that the $z$-axis is the preferred orientation (easy axis). A slight in-plane anisotropy is observed in the $a$-$b$ plane, as indicated by small energy variations along $\phi$, but this effect is much weaker compared to the out-of-plane preference. This behavior marks a significant difference from the bulk counterpart, where the easy axis lies within the plane.

In Figure \ref{Fig:bandsU2}(b), we consider the spin-orbit effects for the AF case (U = 2 eV) 
with projections of the magnetic moments along the z-axis (0, 0, $\pm m_z$), the degeneracy of 
the band structure changes, and we notice a large spin splitting in the valence and conduction bands. 
This effect is particularly noticeable around the high symmetry points M, $\Gamma$, and M', and 
it is larger than in other 2D materials\cite{kosmider2013large}.

\bluemark{The gap opening at the $\Gamma$-point observed in the antiferromagnetic state with SOC (Fig. \ref{Fig:bandsU2}(b)) reflects the interplay between SOC and the symmetry properties of 2D-CRO. Altermagnetic materials inherently break spin-rotational symmetry, which, combined with SOC, reduces the system's symmetry, lifting band degeneracies at high-symmetry points like $\Gamma$. This behavior is consistent with recent studies on altermagnetic systems\cite{smejkal2024altermagnetic}, where it was shown that SOC could lead to symmetry reduction and unconventional splitting patterns. In our system, the observed gap at $\Gamma$ directly results from these symmetry-breaking effects induced by the combined influence of SOC and the antiferromagnetic order.}

Figure \ref{Fig:dosAF} presents the projected density of states (LDOS) for the antiferromagnetic (AF) configuration of the system with $U = 2$ eV. Panel (a) corresponds to the LDOS without spin-orbit coupling (SOC), while panel (b) includes SOC. Comparing both panels, we identify subtle changes in the LDOS, where the Ru $d$-orbitals dominate near the Fermi level, with significant hybridization with the O $p$-orbitals.

The band structure with spin-orbit projected along the direction of the magnetic moments, 
Fig. \ref{Fig:pbands-soc}, we find that the projection along the x-component is symmetric 
to $\Gamma$; therefore, along that direction, the system would behave as a trivial antiferromagnet. 
The projections in the y and z directions show a k-dependent band splitting. For instance, 
following one particular band, the projections change between blue and red as we move from M' to M.
In the projections of the magnetic moment in the y and z directions, we recover the 
alter-magnetic behavior shown for the solution without spin-orbit.

In altermagnetic materials, spin-orbit coupling  breaks the total compensation of spin angular 
momenta, leading to notable ferromagnetic-like behaviors. Specifically, SOC enables the coupling 
between spins and the alternating local structures around the magnetic ions, resulting in non-zero 
net magnetic moments and other ferromagnetic phenomena without external
perturbations\cite{cheong2024altermagnetism}. 
This results in the emergence of unique physical 
phenomena that bridge the characteristics of both ferromagnetic and antiferromagnetic materials. In the next section, we explore other possible phenomena arising from the SOC and AM features.

\subsection{Berry curvature}
\bluemark{Previous studies have investigated the Berry curvature in two-dimensional ruthenate perovskites\cite{groenendijk2020berry,van2021coupling}, particularly \ce{SrRuO3} and Sr$_{0.6}$Ca$_{0.4}$RuO$_3$, across both ferromagnetic and antiferromagnetic phases. These works demonstrated that heterostructure design, interface engineering and strain can control the emergence of topologically non-trivial bands, leading to the observation of the anomalous Hall effect (AHE) in ultra-thin films ruthenate perovskites. These results highlight the potential for manipulating Berry curvature through structural and magnetic phase engineering.}

We calculate the Berry curvature and Chern number using the Fukui's method\cite{fukui2005chern} using the VASPBERRY implementation\cite{Kim_VASPBERRY_2018,kim2022circular}.
The VASPBERRY post-processing tool calculates the the Berry curvature $\Omega_{n}(\mathbf{k})$ for the $n$-th band at a wavevector $\mathbf{k}$.

We use the Berry curvature to calculate the anomalous Hall conductivity $\sigma_{xy}$ integrating the Berry curvature over the Brillouin zone, 
\begin{equation}
    \sigma_{xy} = -\frac{e^2}{\hbar} \sum_{n} \int_{\text{BZ}} \frac{d^2k}{(2\pi)^2} \, f_{n}(\mathbf{k}) \, \Omega_{n}(\mathbf{k}), \nonumber
\end{equation}
where $f_{n}(\mathbf{k})$ is the Fermi-Dirac distribution. The total Hall conductivity is then obtained by summing over the Berry curvature at each $\mathbf{k}$-point and normalizing by the area of the Brillouin zone\cite{gradhand2012first,Kim_VASPBERRY_2018,kim2022circular,nandy2018berry}.

\begin{figure}[t!]
\centering
\includegraphics[clip,width=0.8\columnwidth,angle=0]{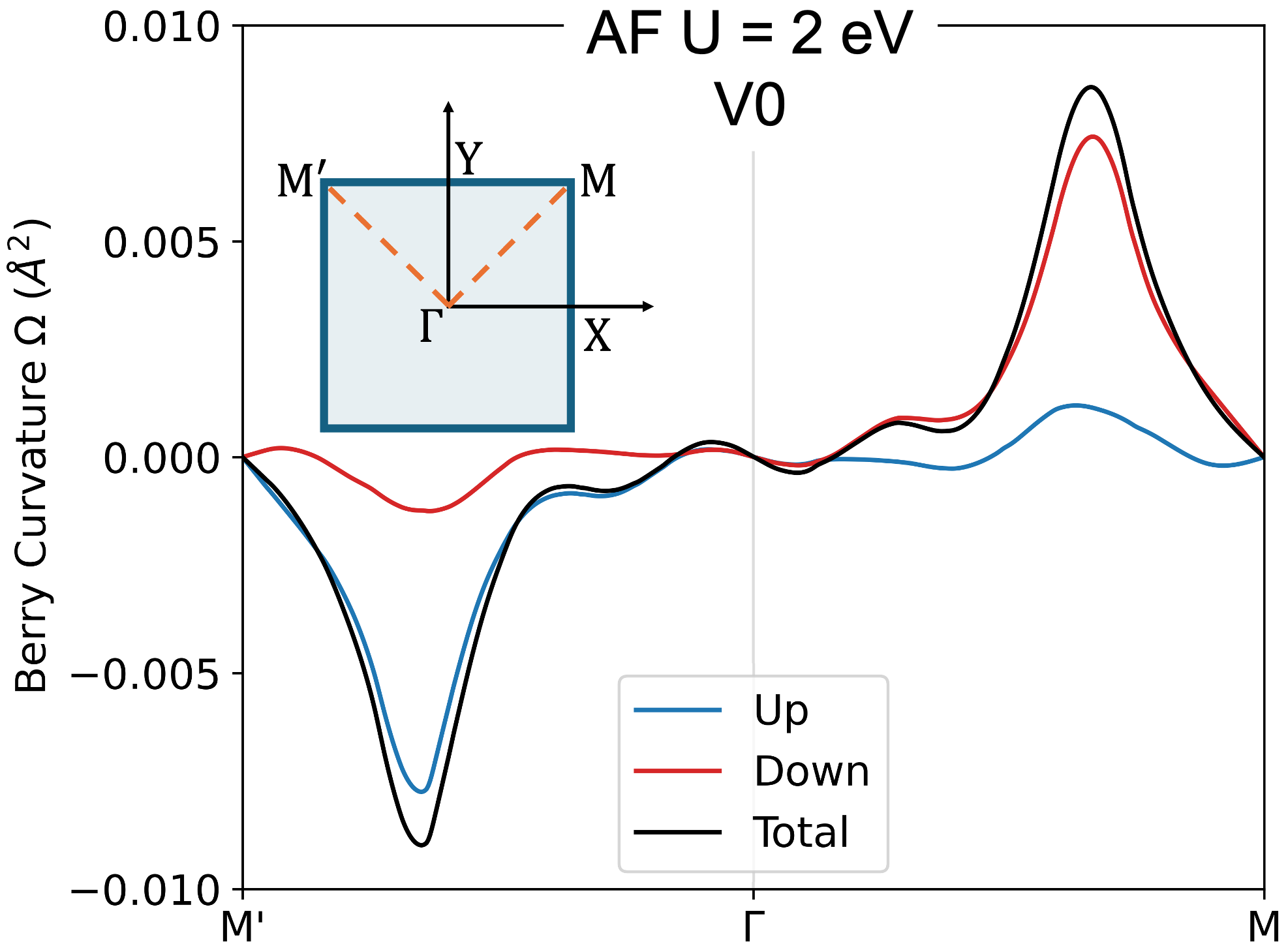} 
\caption{(Color online) Up/Down components and the total of Berry curvature 
along the M'-$\Gamma$-M path for the antiferromagnetic system (U = 2 eV) without 
SOC for the highest occupied valence band (V0). The inset shows the Brillouin zone 
with labeled high symmetry points. 
}
\label{Fig:berry}
\end{figure}

Without spin-orbit coupling, the Berry curvature of the highest occupied valence band (V0) exhibits distinct features that reflect potential non-trivial behavior, as shown in Fig.~\ref{Fig:berry}. Specifically, a negative Berry curvature is observed for spin-up states between the M$'$ and $\Gamma$ points, while a positive Berry curvature appears for spin-down states between $\Gamma$ and M points. Despite these pronounced contributions, the cancellation of the spin-up and spin-down Berry curvature contributions, when integrated over the entire Brillouin zone, yields an overall Chern number that remains $\mathcal{C} = 0$, indicating a trivial topological phase\cite{xiao2010berry}.  
The absence of the spin-orbit effect prevents the breaking of time-reversal symmetry between spin channels, which is critical for realizing non-trivial topological phases such as a quantized Hall conductance \cite{xiao2010berry}. 

\bluemark{To explore potential non-trivial topological phenomena, we investigate the effect of an out-of-plane electric field ($E_z$) on the antiferromagnetic system with $U = 2 \, \text{eV}$, including spin-orbit coupling. 
In Figure \ref{Fig:electric_bands}, we explore the effect of an out-of-plane electric field ($E_z$) on the antiferromagnetic system with $U = 2 \, \text{eV}$ with spin-orbit effects. The application of $E_z$ mainly affects the $e_g$-band. 
In Fig. \ref{Fig:bandsU2}(b), for $E_z = 0.0 \, \text{eV/\AA}$, we follow the doubly degenerate conduction $e_g$-band that at $\text{M}'$ appear around $\sim 2.2 \, \text{eV}$ labeled as C. As $E_z$ increases, this $e_g$-band shifts to lower energies and split for large applied fields, eventually closing the band gap. 
The energy gap is shown to be manipulable with the electric field, as shown in Fig. \ref{Fig:bandsU2}(b) and Fig. \ref{Fig:electric_bands}, for $E_z = 0.0 \, \text{eV/\AA}$, the gap is $0.54 \, \text{eV}$; for $E_z = 0.5 \, \text{eV/\AA}$, it decreases to $0.44 \, \text{eV}$; at $E_z = 1.0 \, \text{eV/\AA}$, the system becomes semi-metallic; and finally, for $E_z = 1.5 \, \text{eV/\AA}$, the band gap is fully closed, showing a metallic character.}

\bluemark{The Berry curvature summed over all occupied states in the $k_x$-$k_y$ plane is sensitive to the applied electric field, $E_z$. The Berry curvature distribution changes in response to the inversion and rotational symmetry breaking induced by the out-of-plane electric field \cite{du2021engineering,abdelrahman2021effects}.}

\bluemark{For low electric fields $E_z \leq 1.0 \, \text{eV/\AA}$, the system behaves as a trivial semiconductor/semimetal with a Chern number $\mathcal{C} = 0$, and the total Hall conductance is negligible ($\sigma_{xy} \sim 10^{-4} \, e^2/h$). This near-zero conductance results from the cancellation of the Berry curvature contributions from occupied states, leading to a topologically trivial phase.}

\bluemark{However, for $E_z = 1.5 \, \text{eV/\AA}$, we observe a significant change in the electronic structure of 2D-CRO, characterized by the closing of the band gap. This transition is accompanied by a topological phase change, where the Chern number changes to $\mathcal{C} = 1$, indicating the onset of a non-trivial topological phase. As a result, the total Hall conductance increases abruptly to $\sigma_{xy} = 4.2 \, e^2/h$, showing the quantized Hall response characteristic of Chern insulators\cite{hu2015chern}.}

\bluemark{This transition to a non-trivial topological phase opens avenues for exploring novel quantum phenomena, such as the quantum anomalous Hall effect (QAHE) \cite{zhang2021spin,gradhand2012first}. In a potential experiment using 2D-CRO within a quantum Hall bar configuration, spin-up and spin-down electrons are predicted to accumulate at opposite edges, consistent with the behavior expected in topological systems. This separation could lead to dissipationless edge currents and robust quantum transport in 2D-CRO, highlighting its potential for applications in topological quantum devices \cite{peng2023intrinsic,chau2021quantum,bal2024quantum}.}

\bluemark{The observed transition to a non-trivial topological phase at $E_z = 1.5 \, \text{eV/\AA}$, aligns with recent experiments measuring the anomalous Hall conductivity in systems with vanishing net magnetization (collinear antiferromagnetic systems) as in the case of Cr-doped \ce{RuO2} \cite{wang2023emergent}, providing experimental evidence of symmetry breaking and the role of Berry curvature in antiferromagnetic transport.}

\bluemark{Band gap engineering in 2D materials often requires the application of large electric fields to achieve significant modifications in the band structure \cite{chaves2020bandgap,haremski2020electrically,tong2016lectures}. 
While an electric field of $E_z = 1.5 \, \text{eV/\AA}$ may seem considerably high, even larger electric fields have been employed in both theoretical and experimental studies on 2D materials. 
For instance, in octagon-nitrogen (ON), an out-of-plane electric field of $2.0 \, \text{eV/\AA}$ was required to close the bandgap and induce a metallic phase \cite{lin2018electronic}. 
Similarly, in Janus MoSSe, a similar electric field is needed to induce a transition from a Schottky to an ohmic contact \cite{nha2024theoretical}.}

\bluemark{Several methods have been proposed to achieve the Quantum Hall Effect (QHE) at lower electric fields, including strain and chemical doping\cite{haremski2020electrically}.
The mechanical strain modifies the band structure and eases QHE-related phenomena by breaking system symmetries and manipulating the band gaps and band crossings\cite{haremski2020electrically}.
Chemical doping introduces extra charge carriers or modifies local electronic environments, enhancing carrier mobility and enabling QHE at lower fields\cite{tong2016lectures}. 
A combined approach using strain, doping, and electric field could be used to achieve QHE at lower electric field conditions. The interplay between strain-induced symmetry breaking, doping-induced carrier enhancement, and electric-field control offers a promising pathway for future device applications. However, further research on controlling the topological character of 2D-CRO is beyond the scope of this paper.}


\section{Toy Model \label{ToyModel}}
\begin{figure}[!]
\centering
\includegraphics[clip,width=.7\columnwidth,angle=0]{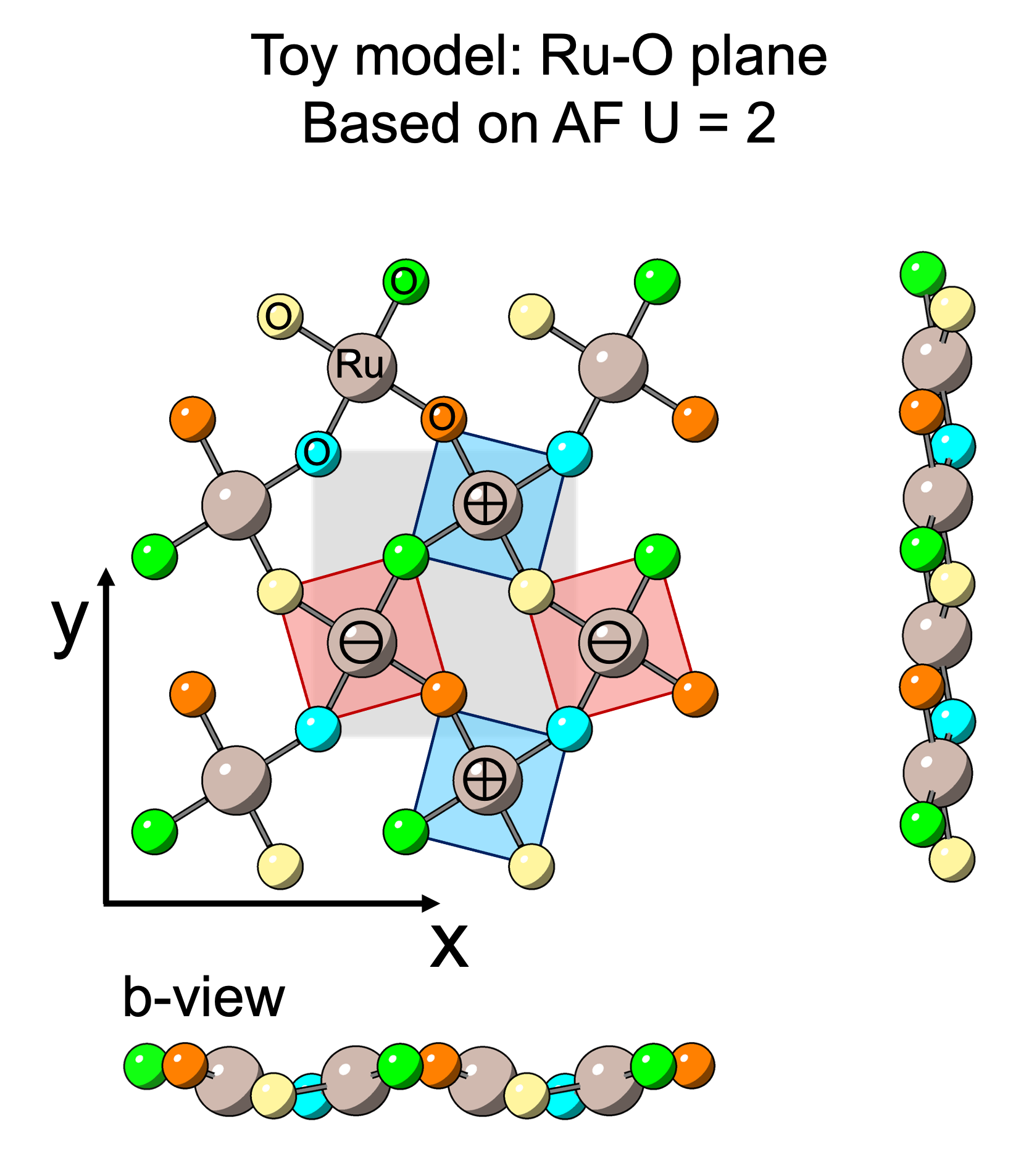} 
\caption{(Color online)  Toy model structure and symmetry operations. We present a ball-and-stick representation of the toy model. The positive and negative projections of the Ru atom magnetic moments are marked as blue and red squares, respectively, while the O atoms are colored based on their relative positions.
The symmetry operations that restore the original magnetic configuration after the time-reversal operation are depicted in Fig. \ref{Fig:toysim}.}
\label{Fig:toy_scheme}
\end{figure}

In order to understand the altermagnetic behavior exhibited by 2D-CRO, we propose a 
toy system that reduces the structure to its essential components, allowing for a 
more transparent analysis of the critical aspects. This model is created by removing 
the top and bottom \ce{Ca-O} layers, leaving the total charge of the unit cell at 
$-2$ e$^-$; as the system is not relaxed, we expect non-zero pressures and forces. 
The toy model consists of two Ruthenium atoms and four staggered Oxygen atoms. 
In the left panel of Figure \ref{Fig:toy_scheme}, we identify two Oxygen atoms 
positioned above the Ruthenium plane (depicted as green and orange spheres); 
the remaining two are located below the plane (cyan and yellow spheres).

In Figure \ref{Fig:toysim}, we show the symmetry operation to recover the original 
configuration after a time-reversal operation (which exchanges the up and down magnetic projections).
Our toy model is the ideal playground to probe possible topological properties in the most 
complex altermagnetic systems.
The toy model geometries are available in the Zenodo repository\cite{Geo}.

From the calculations conducted with the toy model, we conclude that the rotation of the squares defined by the four nearest-neighbor Oxygen atoms around the Ruthenium atoms is the fundamental ingredient for achieving the altermagnetic character of 2D-CRO. To explore the robustness of the altermagnetic character, we attempted to neutralize the charge by adding one hydrogen atom to each Ruthenium atom and aligning the Oxygen atoms in the same plane. \bluemark{Additionally, we modified the aspect ratio between lattice vectors $a$ and $b$. An aspect ratio of $a/b = 1.00$ corresponds to a square base, where the lattice vectors have equal length, while an aspect ratio of $a/b = 0.91$ represents a rectangular base, as observed in the 2D-CRO structure, where $a$ and $b$ differ in length.  }
Despite these changes in lattice constants and the absence of inversion symmetry, the toy model consistently preserved its altermagnetic character. 

\begin{figure}[!]
\centering
\includegraphics[clip,width=1\columnwidth,angle=0]{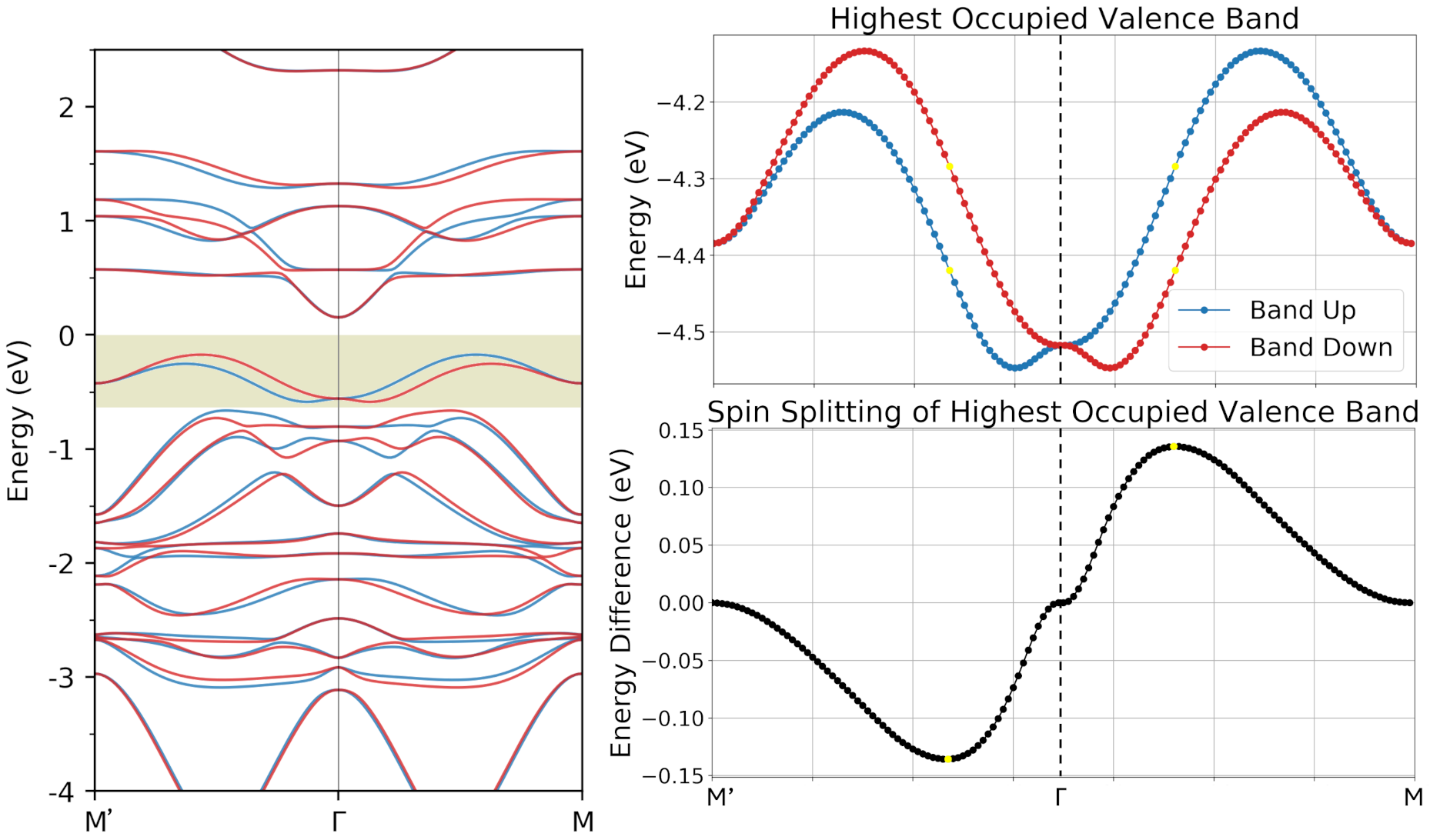} 
\caption{(Color online)  Toy model band structure. In left panel the band structure of the toy model following the same path of Fig. \ref{Fig:bandsU2}. The shaded region corresponds to the highest occupied valence band; in the upper right panel, we show a close-up of that region. The blue and red lines represents the spin-up and spin-down projections. The lower right panel shows the k-dependent splitting corresponding to the energy difference between the up and down bands. We identify the maximum splitting value with a yellow dot in both right panels.
}
\label{Fig:toy_bands}
\end{figure}

Figure \ref{Fig:toy_bands} shows the band structure and the band splitting of the toy model. 
In the left panel of Fig. \ref{Fig:toy_bands}, we plot the band structure; although 
details change concerning the behavior shown by the 2D-CRO (Fig. \ref{Fig:bandsU2}), 
the AM character remains in the M'-$\Gamma$-M path.
For example, if we focus on the highest occupied valence band (the closest to the Fermi level).
When we move away from the high symmetry points, we can see how an up/down band splitting 
appears depending on whether we approach $\Gamma$ from the left (M' side) or the right (M side), 
details in the top-right panel of Fig. \ref{Fig:toy_bands}. For this band, we observe a 
maximum splitting of the valence band is 135.75 meV, which we mark with a yellow dot at right panel. 
As a reference, the maximum splitting of the same band in the 2D-CRO is 218.77 meV (Fig. \ref{Fig:2Dsplitting}).

Note that for a particular band with Ru contribution, the projection of states is mainly 
localized in one of the magnetic atoms, and this contribution is k-dependent. Hence, 
the band structure alternates not only in spin but also in localization.

\section{System Stability}
Once we relax the lattice vectors and atomic positions, we check the stability 
of the system. For this, we use three criteria\cite{gonzalez2024two}: 
\begin{enumerate}
\item The phonon band structure must not have negative frequencies, which indicate imaginary modes. 
\item Molecular dynamics at a fixed temperature above the ambient temperature for a time that allows the thermalization of the system.
\item All the eigenvectors of the Stiffness tensor must be positive. 
\end{enumerate}

The 2D-CRO structure meets all three criteria, which allows us to predict stability in 
a temperature range above the ambient temperature. We checked the system stability, 
focusing on the FM configuration for U = 1 eV and the AF configuration for U = 2 eV.

In Figure \ref{Fig:phonons}, we show the band structure and density of states for the 
phonons of the FM system with U = 1 eV calculated in a $2\times2\times1$  supercell. 
Since the phonon band structure does not present negative values associated with 
imaginary vibrational modes, we can conclude that the structure is dynamically stable. 
Like other perovskite oxides based on Ti and Ta, the maximum frequency is around 
22 THz (100 meV)\cite{gonzalez2024two,ono2023structural}. The calculation of the phonons 
of the AF structure with U = 2 eV leads to a similar behavior already shown in 
Fig. \ref{Fig:phonons} for the FM configuration for U = 1 eV.

Molecular dynamics simulations, as depicted in Figure \ref{Fig:MD}, were conducted 
for the ferromagnetic (U = 1 eV) configuration within a $2\times2\times1$ supercell, 
maintained at a constant temperature of 600 K over though a N\'ose-Hover thermostat 
a time of 10 ps. The MD simulations demonstrate the stability of the system in a 
range of temperatures above room temperature.

Finally, the elastic properties of 2D materials are evaluated within the harmonic 
range of elastic deformations ($|\varepsilon| < 2.5\,\% $) using the finite 
difference method and strain-stress implemented in the VASPKIT code\cite{wang2021vaspkit}. 
We introduce small in-plane deformations on the lattice vectors and study the 
energy/strain of the resulting relaxed geometries\cite{mouhat2014necessary,zhou2013theoretical}.

The stiffness tensor in N/m for 2D-CRO in the FM configuration (U = 1 eV) is,
\begin{equation}
\mathbf{C}_{FM} = 
\begin{bmatrix}
     44.37  &   11.89 &     0.00 \\
     11.89  &   44.37 &     0.00 \\
      0.00 &    0.00 &    85.55 \\
\end{bmatrix}, \nonumber
\end{equation}
similarly, for the AF configuration (U = 2 eV) the stiffness tensor reads,
\begin{equation}
\mathbf{C}_{AF} = 
\begin{bmatrix}
     37.41   &  36.46  &    0.00 \\
     36.46   &  50.18  &    0.00\\
      0.00   &   0.00  &   54.05 \\
\end{bmatrix}. \nonumber
\end{equation}
We confirm that the eigenvalues of the stiffness matrix in both cases are positive, 
and the conditions $C_{11}>0$, $C_{66}>0$, and $C_{11}C_{22}  >  C_{12} C_{12} $ 
are fulfilled. These results satisfy the established 
criteria\cite{mouhat2014necessary, wang2022high} and allow us to conclude that 
regardless of the magnetic character, our 2D-CRO monolayer exhibits mechanical stability.

A detailed report of the elastic properties of 2D-CRO in the FM (U = 1 eV) and AF (U = 2 eV) 
configurations are included in Fig. \ref{Fig:elastic}. The marked differences in the 
elastic properties between FM and AF configurations, especially in Young's and 
shear modulus, suggest the influence of super-exchange interactions between 
Ru atoms and their in-plane O neighbors\cite{wang2023recent}.

\section{Final remarks}
In summary, we have used density functional theory to explore the elastic, magnetic, and electronic properties of the \ce{Ca2RuO4} monolayer. Using three independent criteria, we predict the thermal and mechanical stability of the 2D-CRO layer. Consistent with observations in bulk, we identified a simultaneous change in the system's geometry and magnetism associated with electron correlation effects. Subtle changes in the Ru-O in-plane and out-of-plane distances, rotations, and tilts around the magnetic atoms stabilize antiferromagnetic solutions, leading to the emergence of altermagnetism. The study of Berry curvature and the Chern number indicates the potential for topological states with and without spin-orbit coupling effects. 
\bluemark{The spin-orbit effects and external electric field ($E_z$) could lead to a topological phase change, as evidenced by the evolution of the Berry curvature and the Chern number. At large electric fields, the system transitions from a trivial insulator ($C=0$) to a Chern insulator with a non-trivial topology ($C=1$).}
Our findings suggest that Ruddlesden-Popper materials are promising candidates for exploring and manipulating altermagnetic properties and non-trivial topological phases at lower dimensionalities.
 
Finally, we propose a toy model that allows us to explore the effects of geometry and the role of inversion symmetry in two-dimensional altermagnetic systems. The proposed toy model is an ideal playground for studying these factors and can be easily applied to more complex altermagnetic systems.

\section*{Acknowledgments}
The authors would like to acknowledge the financial support from Chilean FONDECYT grants numbers 11220854, 1210607, 1221301 and 1220700.
AL thanks to Fondecyt Postdoctoral Grant No. 3220505 and to the Dresden Fellowship Grant.

\section*{Competing Interests}

The Authors declare no Competing Financial or Non-Financial Interests.

\section*{Data Availability}

The data that support the findings of this study are available from the corresponding author, upon reasonable request.


\pagebreak
\newpage\null\thispagestyle{empty}\newpage
\widetext

\begin{center}
\textbf{\large Supplemental Material: \\ Altermagnetism in Two Dimensional \ce{Ca-Ru-O} Perovskite}

J. W. González,$^{1,*}$ A.M León,$^{2,3}$ C. González-Fuentes,$^{4}$ and R. A. Gallardo$^{5}$ 

$^1$ Departamento de Física, Universidad de Antofagasta, Av. Angamos 601, Casilla 170, Antofagasta, Chile.\\
$^2$ Departamento de Física, Facultad de Ciencias, Universidad de Chile, Casilla 653, Santiago, Chile.\\ 
$^3$ Institute for Solid State and Materials Physics, TU Dresden University of Technology, 01062 Dresden, Germany. \\
$^4$ Instituto de Física, Pontificia Universidad Católica de Chile, Vicuña Mackena 4860, 7820436 Santiago, 
Chile. \\
$^5$ Departamento de Física, Universidad Técnica Federico Santa María, Avenida España 1680, Valparaíso, Chile.\\
$^{*}$ Electronic address: jhon.gonzalez@uatonf.cl\\
\end{center}

\setcounter{figure}{0} 
\setcounter{section}{0} 
\setcounter{equation}{0}
\setcounter{page}{1}
\renewcommand{\thepage}{S\arabic{page}} 
\renewcommand{\thesection}{S\Roman{section}}   
\renewcommand{\thetable}{S\arabic{table}}  
\renewcommand{\thefigure}{S\arabic{figure}} 
\renewcommand{\theequation}{S\arabic{equation}} 

\section{Relation between Hubbard-U term and Geometry}
Relationship between geometrical parameters and magnetic moment in Ruthenium atoms as a function of the Hubbard-U parameter, Table \ref{Tab:a}.

\begin{table}[h!]
    \centering
    \begin{tabular}{|c||c|c|c|c|c|}
    \hline
Magnetic & a & a/b & Ru  & O in-plane & O out-plane  \\ 
conf. & (\AA{}) &   &  mag. ($\mu_B$) & mag. ($\mu_B$) & mag. ($\mu_B$) \\  \hline \hline
        \textbf{FM} (U$= 0$ eV) & 5.136 & 1.03 & 0.93 & 0.13 & 0.01 \\ \hline
        AF (U$= 0$ eV)  & 5.167 & 1.00 & 0.71 & 0.00 & 0.01 \\ \hline \hline
        \textbf{FM} (U$= 1$ eV) & 5.200 & 1.00 & 1.44 & 0.19 & 0.02 \\ \hline
        AF (U$= 1$ eV)  & 5.168 & 0.99 & 1.15 & 0.07 & 0.02 \\ \hline \hline \hline
        FM (U$= 2$ eV) & 5.169 & 0.99 & 1.48 & 0.18 & 0.02 \\ \hline
        \textbf{AF} (U$= 2$ eV) & 5.174 & 0.91 & 1.46 & 0.01 & 0.12 \\ \hline \hline
        FM (U$= 3$ eV) & 5.174 & 0.98 & 1.55 & 0.15 & 0.02 \\ \hline
        \textbf{AF} (U$= 3$ eV) & 5.196 & 0.91 & 1.53 & 0.01 & 0.11 \\ \hline
    \end{tabular}
\caption{Summary with the lattice constant (a), aspect ratio (a/b), Ruthenium magnetic moment, 
and Oxygen (in-plane and out-plane) magnetic moments as a function of the Hubbard-U term. 
The boldface labels indicate the ground state of each U value.}\label{Tab:a}   
\end{table}

\begin{figure}[h!]
\centering
\includegraphics[clip,width=.45\columnwidth,angle=0]{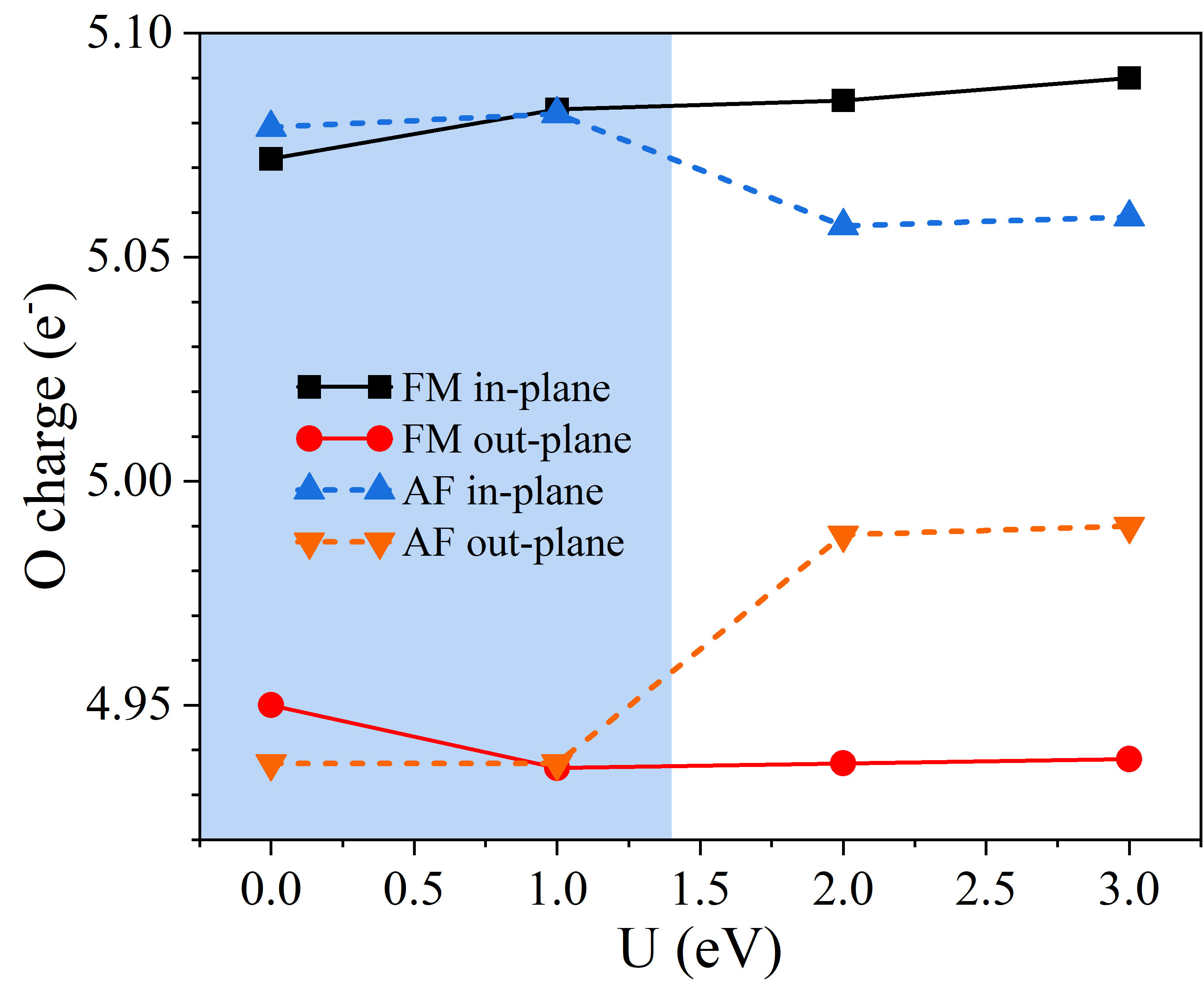} 
\caption{Bader charges analysis for oxygen atoms (in $e^-$) as a function of the Hubbard $U$ parameter for ferromagnetic (FM) and antiferromagnetic (AF) configurations, with both in-plane and out-of-plane magnetic orientations.}
\label{Fig:bader}
\end{figure}

Bader charges analysis for oxygen atoms in Figure \ref{Fig:bader} data show that the oxygen charge slightly increases with $U$ for FM configurations, with in-plane orientation exhibiting higher values than out-of-plane. In the AF case, the oxygen charge remains nearly constant for in-plane orientation but shows a distinct increase for the out-of-plane orientation as $U$ increases. The shaded region indicates the range of $U$ values used in the main calculations.

\section{Ising model \label{sec:ising}}

In the two-dimensional Ising model used in this analysis, we consider periodic boundary conditions in a  $30\times30\time1$ supercell with a nearest-neighbor interaction described by the exchange constant $J$. $J > 0$ for ferromagnetic interaction, while $J < 0$ indicates antiferromagnetic interactions.

The energy of the system is given by:
\[
E = -J \sum_{\langle i,j\rangle} \sigma_{i}\sigma_{j},
\]
where the sum is taken over all pairs of neighboring spins $\langle i, j\rangle$ and each lattice site is a spin variable $\sigma_{i,j} = \pm 1$

\begin{figure}[h!]
\centering
\includegraphics[clip,width=.8\columnwidth,angle=0]{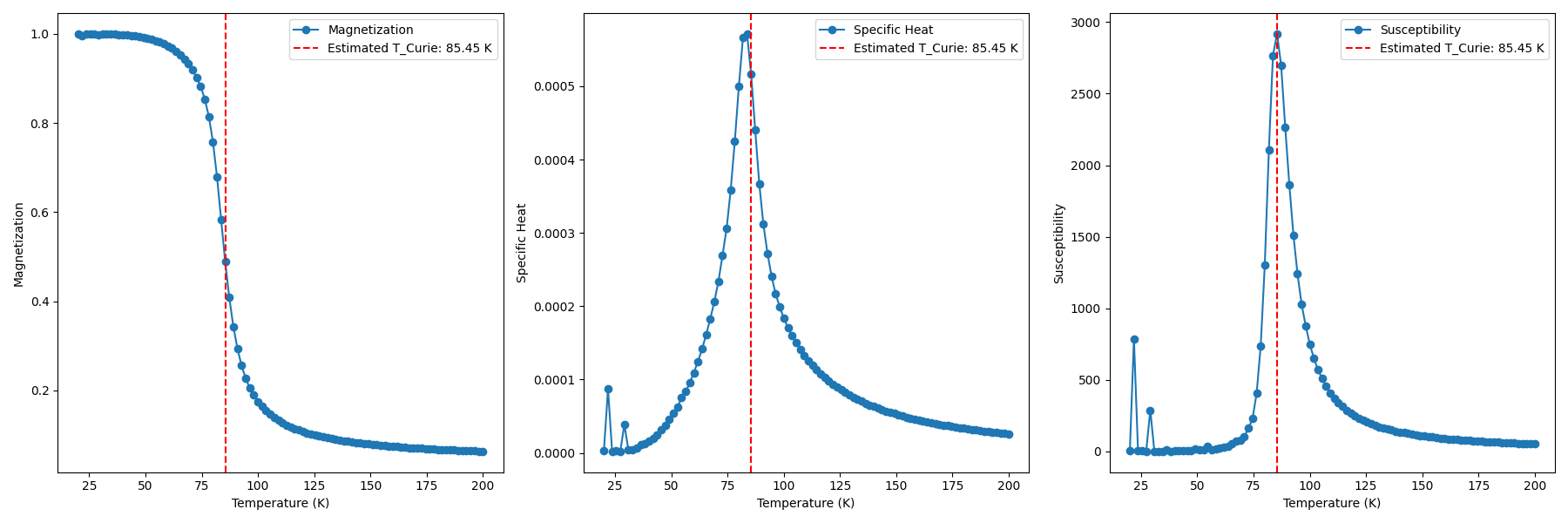} 
\caption{(Color online) Ising model for an \textbf{FM system} with $J = 3.1$ meV: Magnetization, specific heat and susceptibility as a function of temperature.}
\label{Fig:isingFM_J03}
\end{figure}

\begin{figure}[h!]
\centering
\includegraphics[clip,width=.85\columnwidth,angle=0]{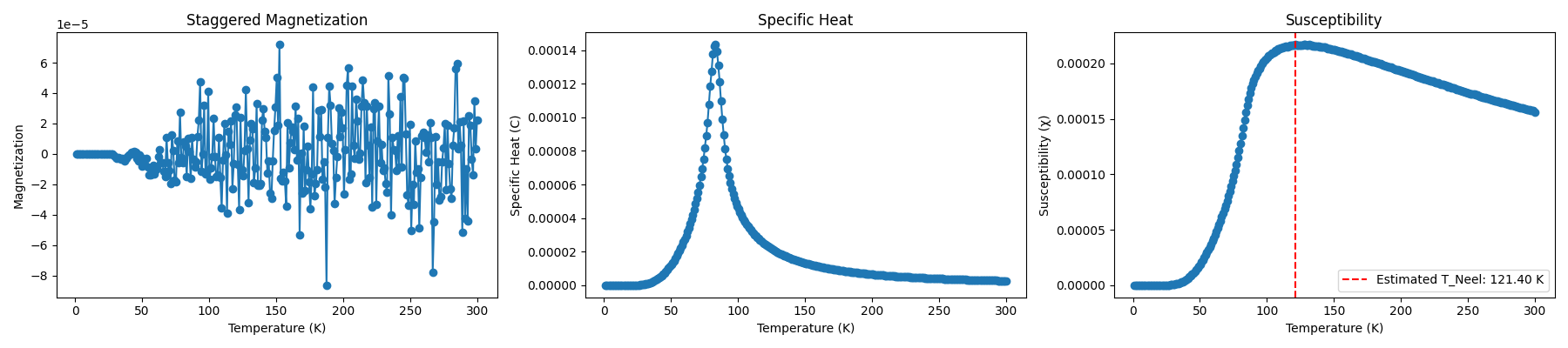} 
\caption{(Color online) Ising model for an \textbf{AF system} with $J = -3.1$ meV: Magnetization, specific heat and susceptibility as a function of temperature.}
\label{Fig:isingAF_J03}
\end{figure}

We use Monte Carlo simulations using the Metropolis algorithm used to estimate physical quantities over a range of temperatures:
\begin{itemize}
    \item A random spin is selected and flipped with a probability dependent on the change in energy, $\Delta E$, given by:
    \[
    \Delta E = -2J \sigma_{i,j}\left(\sigma_{i+1,j} + \sigma_{i-1,j} + \sigma_{i,j+1} + \sigma_{i,j-1}\right),
    \]
    which represents the energy difference between the initial and proposed states.
    \item If $\Delta E < 0$, the spin is flipped unconditionally. Otherwise, the spin is flipped with probability:
    \[
    P = \exp\left(-\frac{\Delta E}{k_B T}\right),
    \]
    where $k_B$ is the Boltzmann constant and $T$ is the absolute temperature.
\end{itemize}

For each temperature, we let the system thermalize for 500 steps, and we take the average over the following 2000 simulation steps; later, we calculate the following observables:

\textbf{Magnetization}: Average absolute value of the total spin, normalized by the number of lattice sites,
    \[
    M = \frac{\left|\sum_{i,j}\sigma_{i,j}\right|}{L^2},
    \]
    where $L$ is the lattice size.

\textbf{Specific Heat}: Derived from the energy fluctuations:
    \[
    C = \frac{\langle E^2 \rangle - \langle E \rangle^2}{k_B T^2 L^2}.
    \]

\textbf{Susceptibility}: Derived from the magnetization fluctuations:
    \[
    \chi = \frac{\langle M^2 \rangle - \langle M \rangle^2}{k_B T L^2}.
    \]

The Curie and Neél temperatures represent the critical temperatures at which ferromagnetic and antiferromagnetic materials undergo a phase transition to a paramagnetic phase, respectively.

On the one hand, in ferromagnetic systems, the Curie temperature is estimated from the temperature at which the magnetic susceptibility peaks. In Figure \ref{Fig:isingFM_J03}, we present averaged properties for the FM system. For $J = 3.10$ meV, the $T_C = 85.45$ K is closer to the reported values for similar systems\cite{nobukane2020co}.

On the other hand, in antiferromagnetic configuration, the Neél temperature is also estimated using the peak of the magnetic susceptibility. The AF calculations for $J= -3.10$ meV yields to $T_N = 121.4$ K; details are shown in Fig. \ref{Fig:isingAF_J03}.

\section{AF band splitting}
\begin{figure}[h!]
\centering
\includegraphics[clip,width=.45\columnwidth,angle=0]{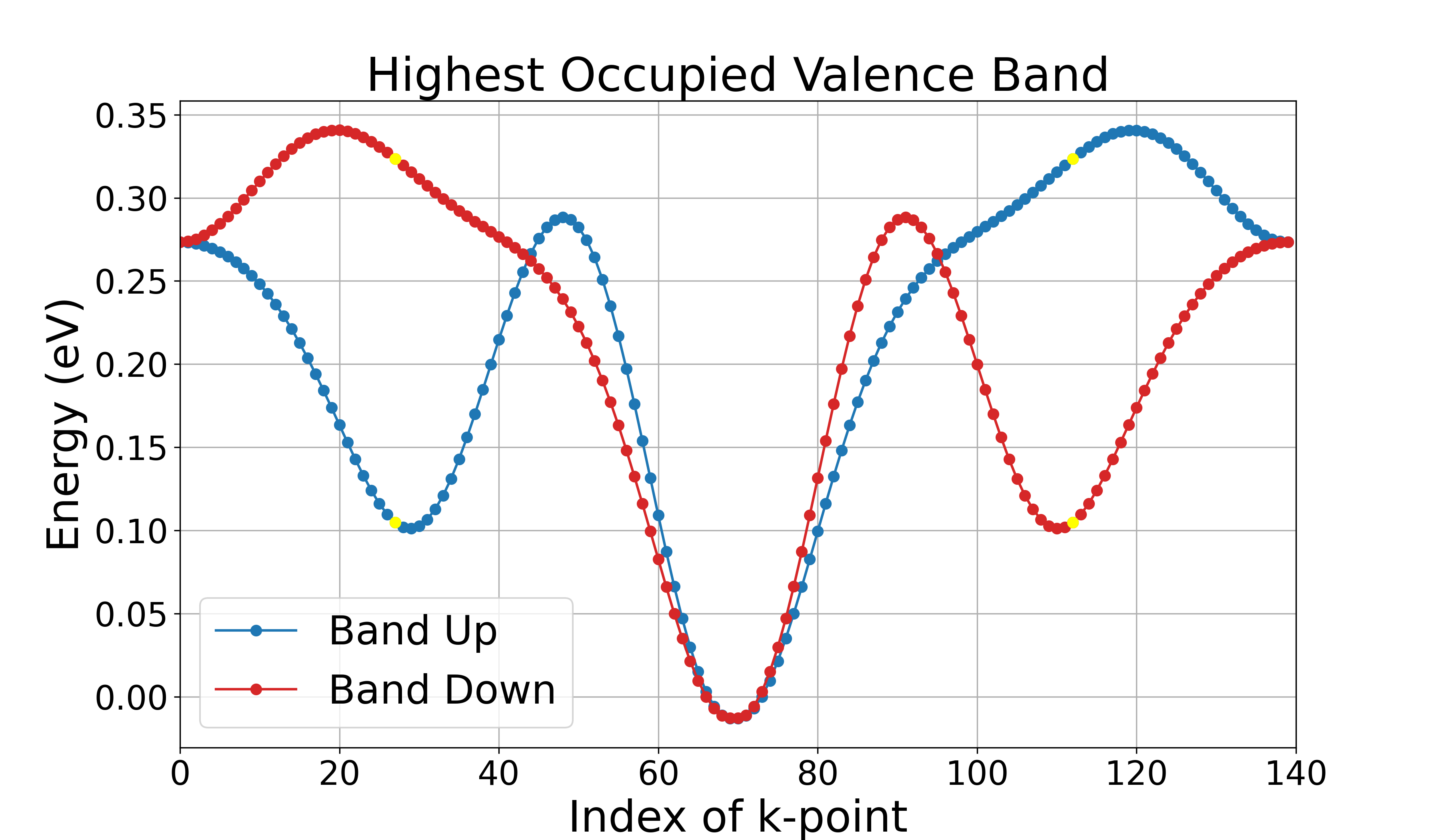} 
\includegraphics[clip,width=.45\columnwidth,angle=0]{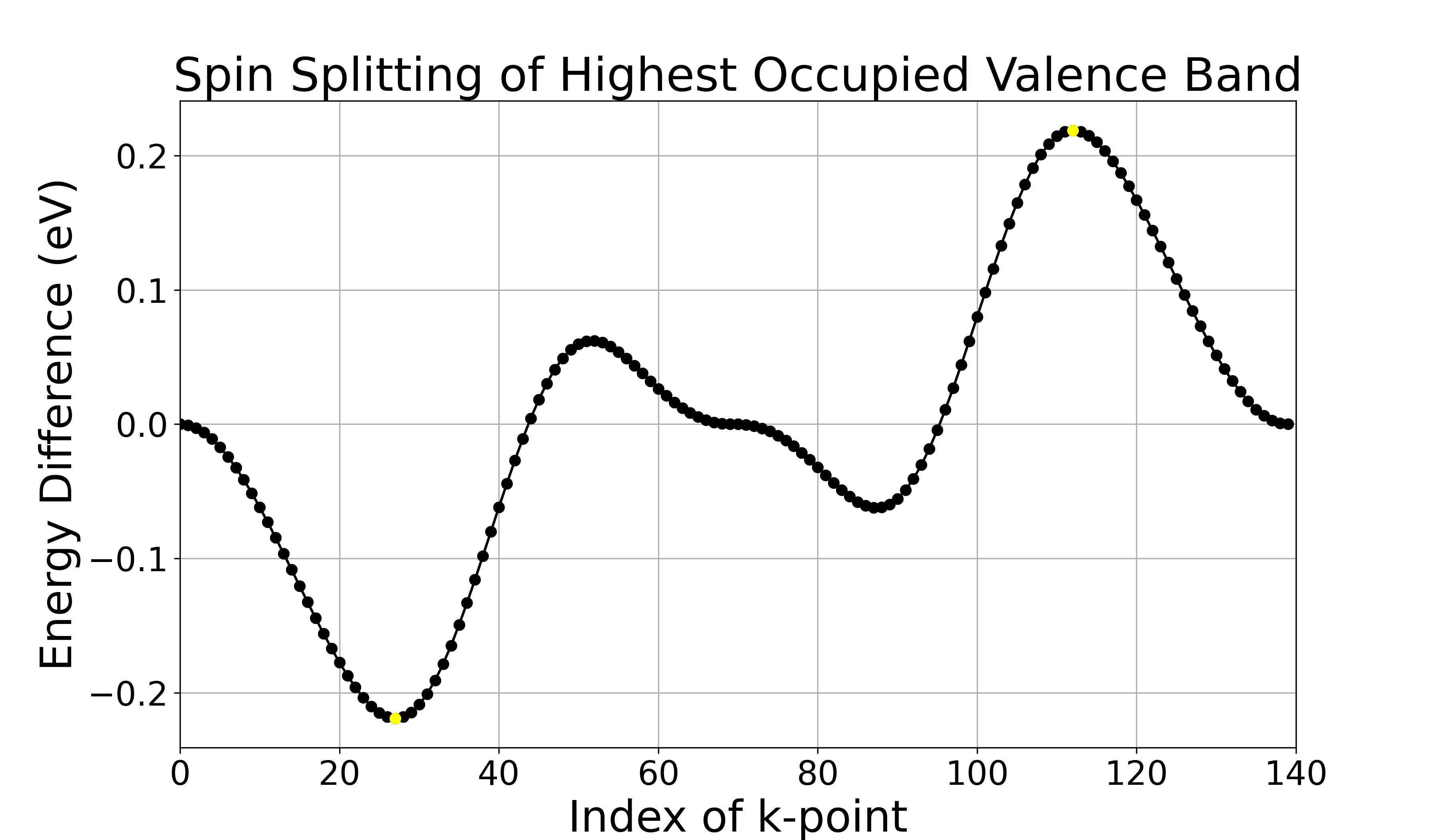} 
\caption{(Color online) Detail of the highest occupied valence band V0 (without spin-orbit effects) along the M'-$\Gamma$-M path for the antiferromagnetic (U = 2 eV) system. }
\label{Fig:2Dsplitting}
\end{figure}

In figure \ref{Fig:2Dsplitting} we present the details of the highest occupied valence band, the left panel, correspond to the zoom of the band structure shown in Fig. \ref{Fig:bandsU2} (a). The right panel of Fig. \ref{Fig:2Dsplitting} shows the energy difference between the up and down bands.

\section{Local density of states \label{sec:ldos}}
\begin{figure}[!]
\centering
\includegraphics[clip,width=0.45\columnwidth,angle=0]{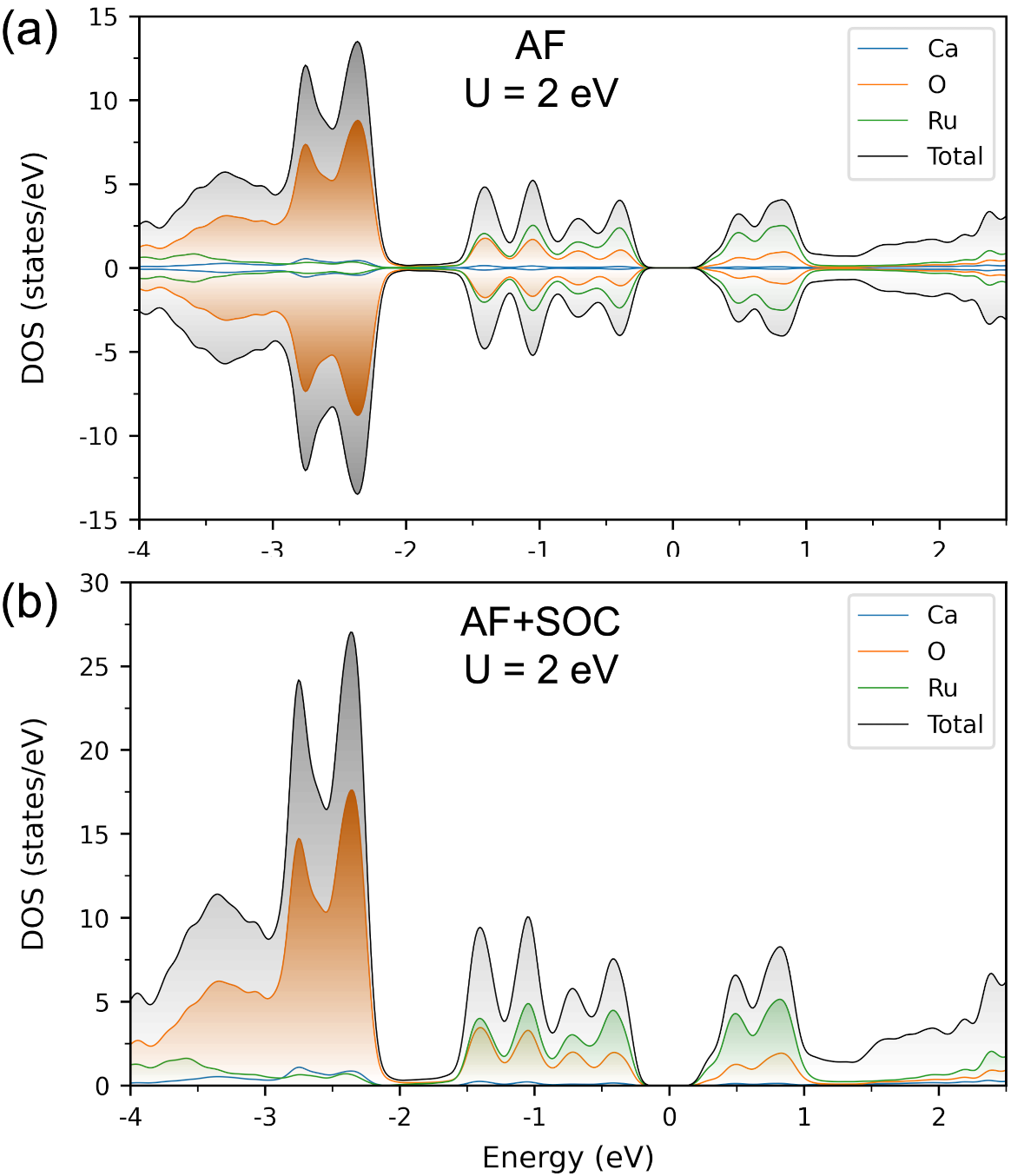} 
\caption{(Color online) The density of states for the antiferromagnetic system with U = 2 eV. In (a) without spin-orbit and (b) with spin-orbit coupling.
The colors represent the contribution of each atom. 
}
\label{Fig:dosAF}
\end{figure}

The local density of states for the antiferromagnetic case with U = 2 eV, in Figure \ref{Fig:dosAF} shows that with and without spin-orbit coupling, the system is a semiconductor. The states near the Fermi level show a high contribution of orbitals from the Ruthenium and Oxygen atoms, favoring super-exchange interactions between the d-Ru and p-O orbitals.

\begin{figure}[!]
\centering
\includegraphics[clip,width=0.49\columnwidth,angle=0]{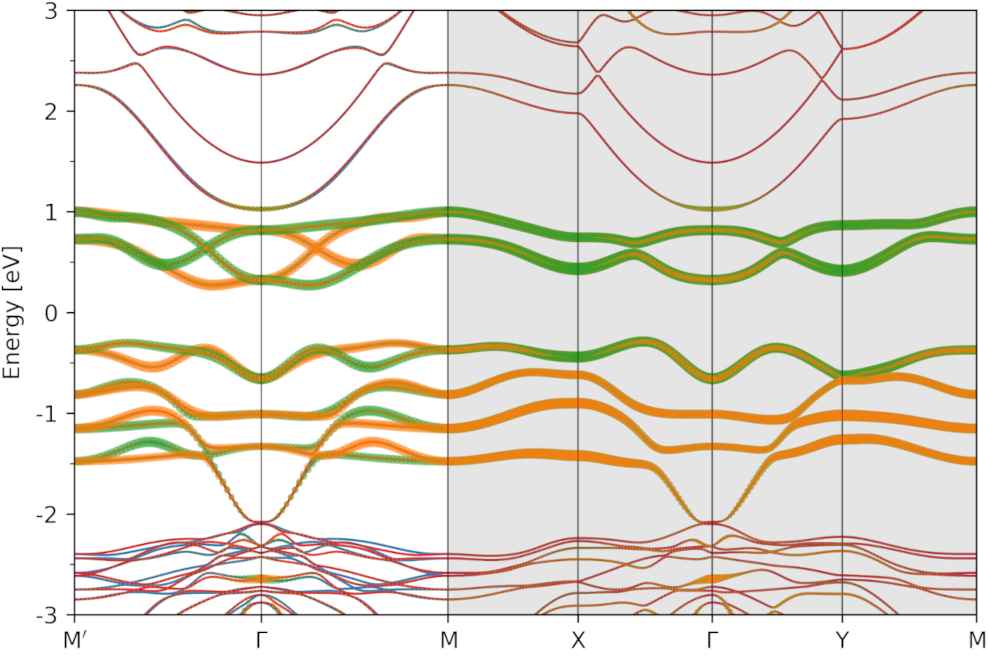}
\includegraphics[clip,width=0.49\columnwidth,angle=0]{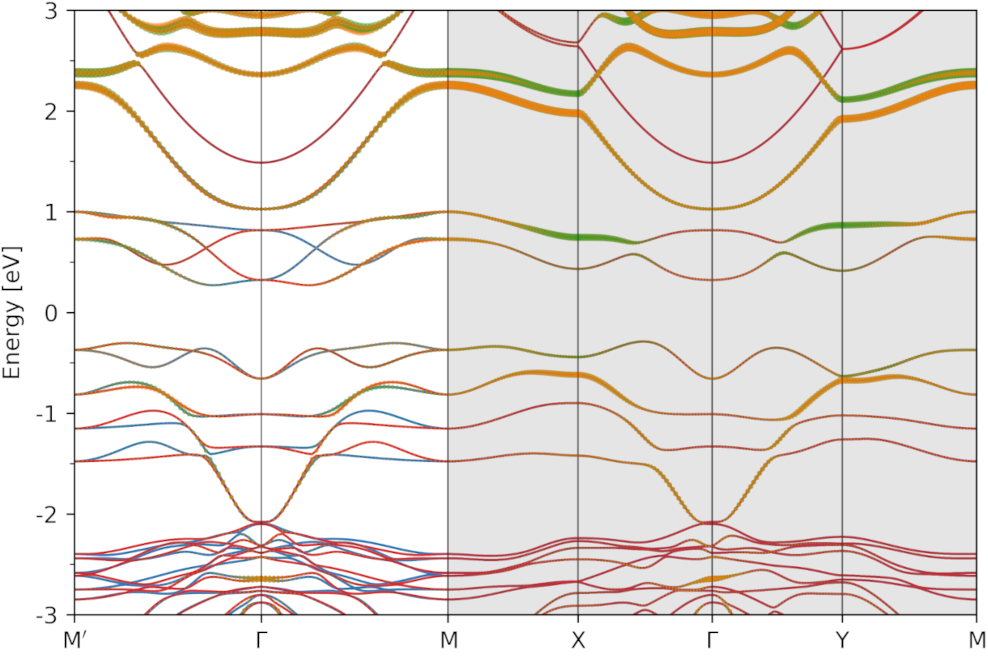} 
\caption{Projected band structure for the antiferromagnetic system with $U = 2 \, \text{eV}$.  The left panel shows the $t_{2g}$ orbital projection, while the right panel shows the $e_g$ orbital projection.  The orange and green colors represent the contributions of different Ru atoms.  The energy axis is referenced to the Fermi level at $E = 0 \, \text{eV}$.
Note that this band structure corresponds to an extended path of Fig.~\ref{Fig:bandsU1}(a) where the system only presents altermagnetic character along the M'$-\Gamma-$M path.}
\label{Fig:pband_AF}
\end{figure}

\section{Non-collinear band structure \label{sec:pband}}
\begin{figure}[h!]
\centering
\includegraphics[clip,width=.9\columnwidth,angle=0]{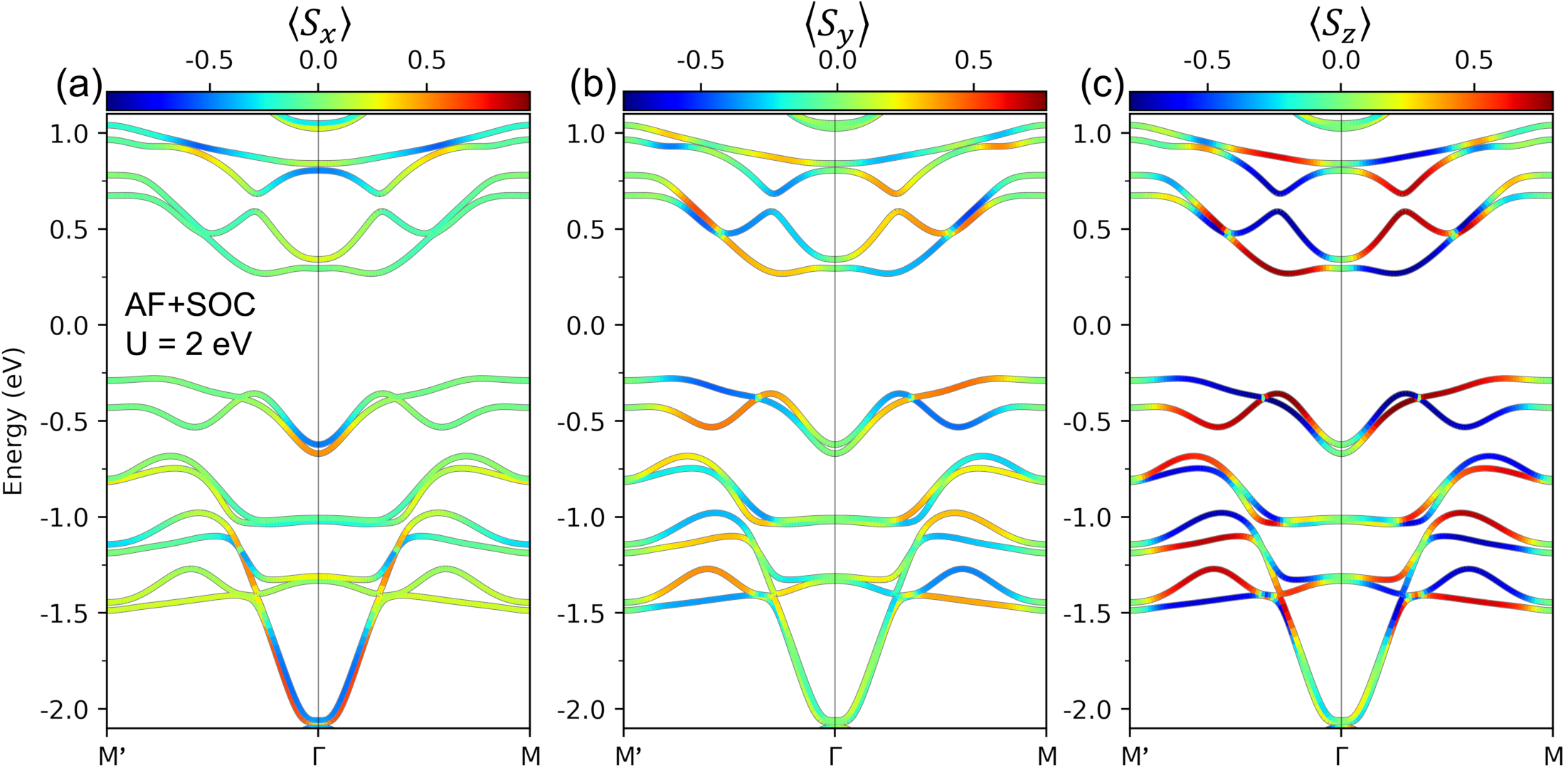} 
\caption{(Color online) Projected band structure along the M'-$\Gamma$-M path for the antiferromagnetic (U = 2 eV) system with spin-orbit coupling. The magnetic moments are projected along the (a) x-axis, (b) y-axis, and (c) z-axis. The color scale represents the expectation value of the spin component, with blue indicating negative values, red indicating positive values. Note the k-dependent band splitting and the differences in spin projection along each axis.
}
\label{Fig:pbands-soc}
\end{figure}

In Fig. \ref{Fig:pbands-soc} we present the band structure along the M'-$\Gamma$-M path for the antiferromagnetic system (U = 2 eV) with spin-orbit coupling, with magnetic moments projected along the x-axis (a), y-axis (b), and z-axis (c).

\section{Electric field  \label{sec:electric}}
In this section, we explore the effect of an out-of-plane electric field ($E_z$) on the band structure and Berry curvature for the antiferromagnetic configuration ($U = 2 \, \text{eV}$) with spin-orbit coupling. We analyze the system for various electric field strengths: $E_z = 0.0 \, \text{eV/\AA}$, $E_z = 0.5 \, \text{eV/\AA}$, $E_z = 1.0 \, \text{eV/\AA}$, and $E_z = 1.5 \, \text{eV/\AA}$.

\begin{figure}[h!]
\centering
\includegraphics[clip,width=.9\columnwidth,angle=0]{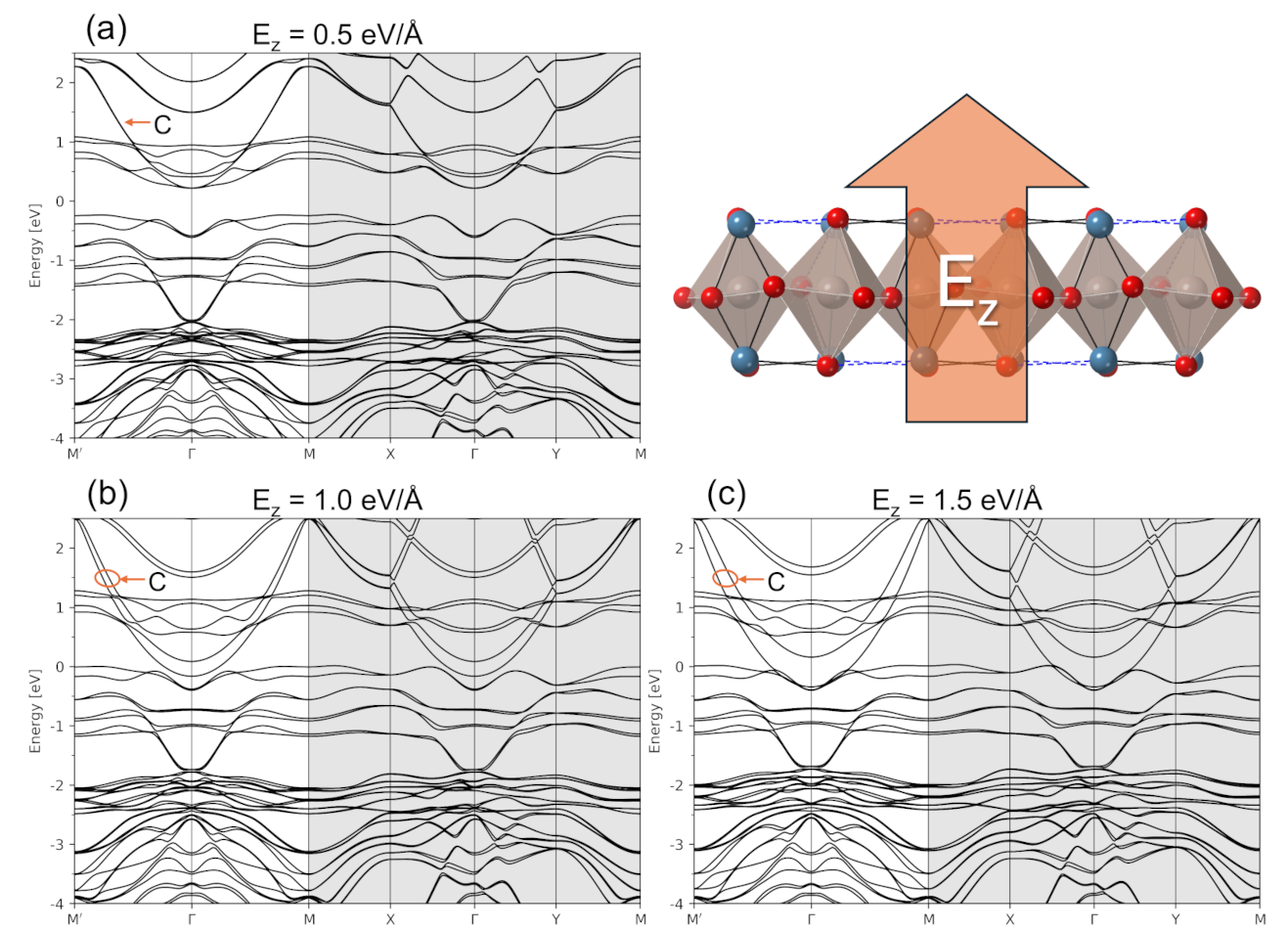} 
\caption{Band structure for the antiferromagnetic configuration ($U = 2 \, \text{eV}$) with spin-orbit coupling under an out-of-plane electric field ($E_z$).  Panel (a) correspond to $E_z = 0.5 \, \text{eV/\AA}$, (b) $E_z = 1.0 \, \text{eV/\AA}$, and (c) $E_z = 1.5 \, \text{eV/\AA}$.  The band structure for $E_z = 0.0 \, \text{eV/\AA}$ is shown in Fig.~\ref{Fig:bandsU2}(b). 
The schematic shows the direction of the applied electric field relative to the 2D lattice.  }
\label{Fig:electric_bands}
\end{figure}

The band labeled 'C' is an $e_g$-band presents a higher sensibility to electric field. As $E_z$ increases, this band stretches progressively to lower energies, eventually closing the energy gap near the $\Gamma$ point. Figure~\ref{Fig:electric_bands} shows the evolution of the band structure under different $E_z$ values, while Figure~\ref{Fig:bandsU2}(b) shows the reference band structure for $E_z = 0.0 \, \text{eV/\AA}$.

\begin{figure}[h!]
\centering
\includegraphics[clip,width=.9\columnwidth,angle=0]{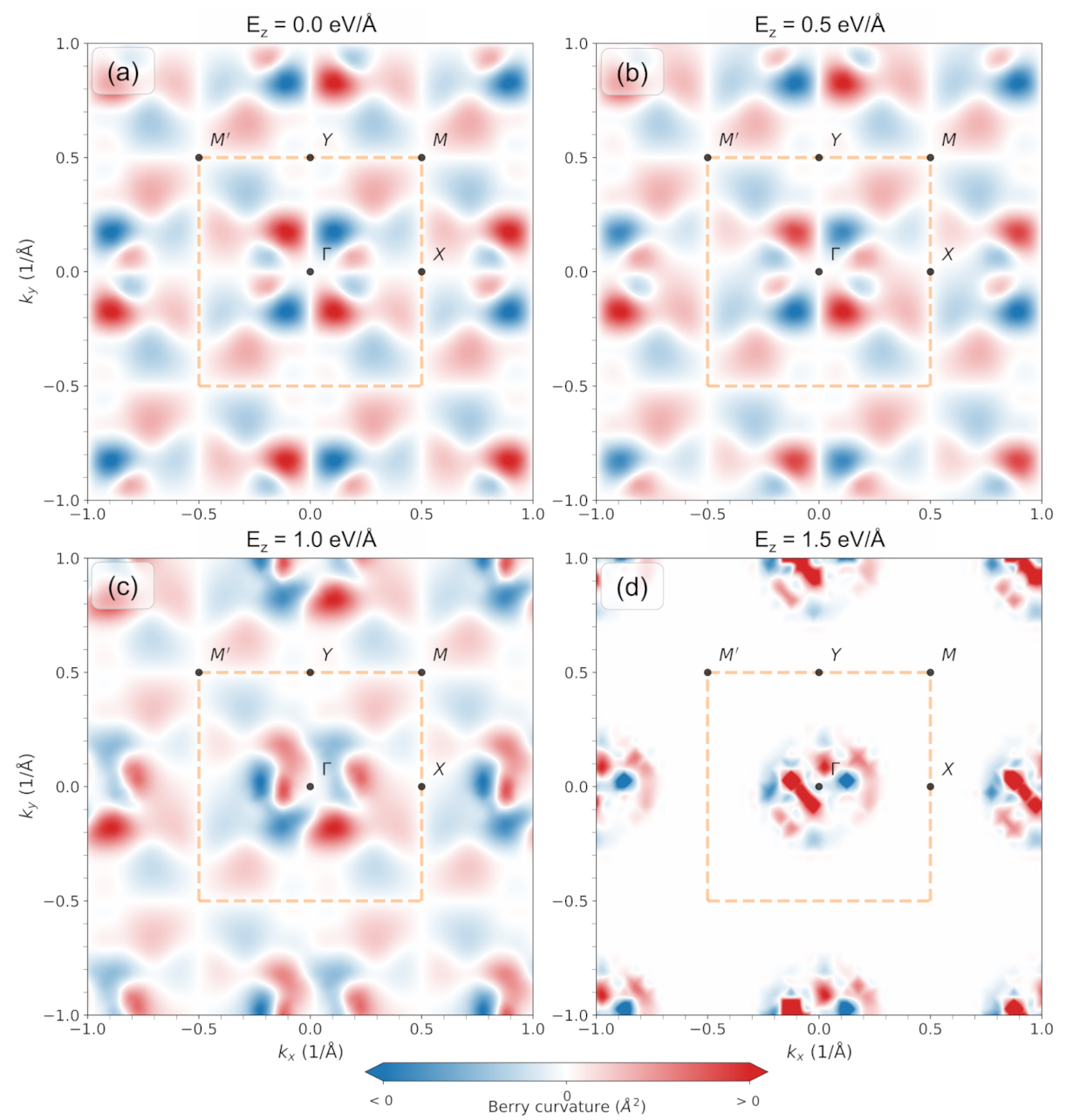} 
\caption{Berry curvature summed over all occupied states in the $k_x$-$k_y$ plane for the antiferromagnetic configuration ($U = 2 \, \text{eV}$) with spin-orbit coupling for out-of-plane electric fields ($E_z$).  Panel (a) correspond to $E_z = 0.0 \, \text{eV/\AA}$, (b) $E_z = 0.5 \, \text{eV/\AA}$, (c) $E_z = 1.0 \, \text{eV/\AA}$, and (d) $E_z = 1.5 \, \text{eV/\AA}$.
The dashed orange box outlines the first Brillouin zone, highlighting the high-symmetry points: $\Gamma$, X, M, Y, and M$'$. The blue and red areas represent negative and positive values of Berry curvature, respectively. 
}
\label{Fig:electric_berry}
\end{figure}

Berry curvature summed over all occupied states in the $k_x$-$k_y$ plane, in Figure~\ref{Fig:electric_berry}, shows a clear evolution with increasing out-of-plane electric field ($E_z$). For $E_z = 0.0 \, \text{eV/\AA}$, the Berry curvature distribution is symmetric with alternating positive and negative regions. As $E_z$ increases, this symmetry is progressively broken, leading to more pronounced and localized regions of Berry curvature, particularly around the $\Gamma$ point at $E_z = 1.5 \, \text{eV/\AA}$. 

\newpage

\section{Magnetic anisotropy energy (MAE) \label{sec:MAE}}
\begin{figure}[h!]
\centering
\includegraphics[clip,width=.6\columnwidth,angle=0]{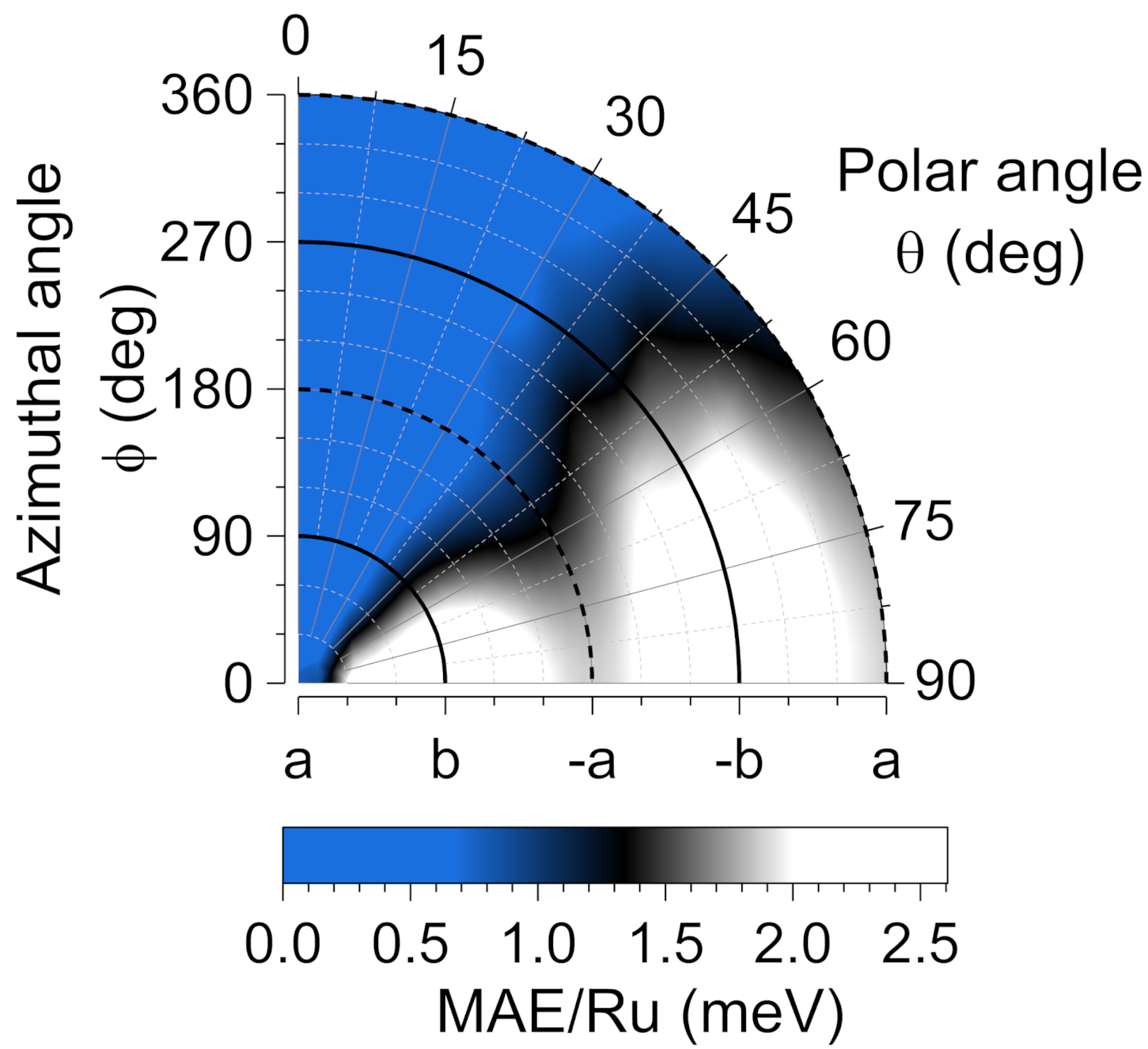} 
\caption{(Color online) Magnetic anisotropy energy (MAE) of the ferromagnetic (FM) system (U = 1 eV) with spin-orbit coupling effects. The contour plot of the energy difference ($E(\theta, \phi)-E_{GS}$) per Ru atom as a function of the azimuthal angle ($\phi$) and polar angle ($\theta$) of the magnetic moment direction. The color scale represents the MAE, with blue indicating lower energy and white indicating higher energy. The ground state, corresponding to the magnetic moment pointing in the z-direction, is identified as the easy axis, and the system shows a small in-plane anisotropy.
}
\label{Fig:MAE}
\end{figure}

After calculating the energy difference for the FM system (with spin-orbit effects) with the magnetic moment pointing in several directions, we have identified the easy axis along the z-direction and small in-plane anisotropy. Details in Figure \ref{Fig:MAE}.

\newpage

\section{Toy model \label{sec:toy}}

\begin{figure}[h!]
\centering
\includegraphics[clip,width=0.8\columnwidth,angle=0]{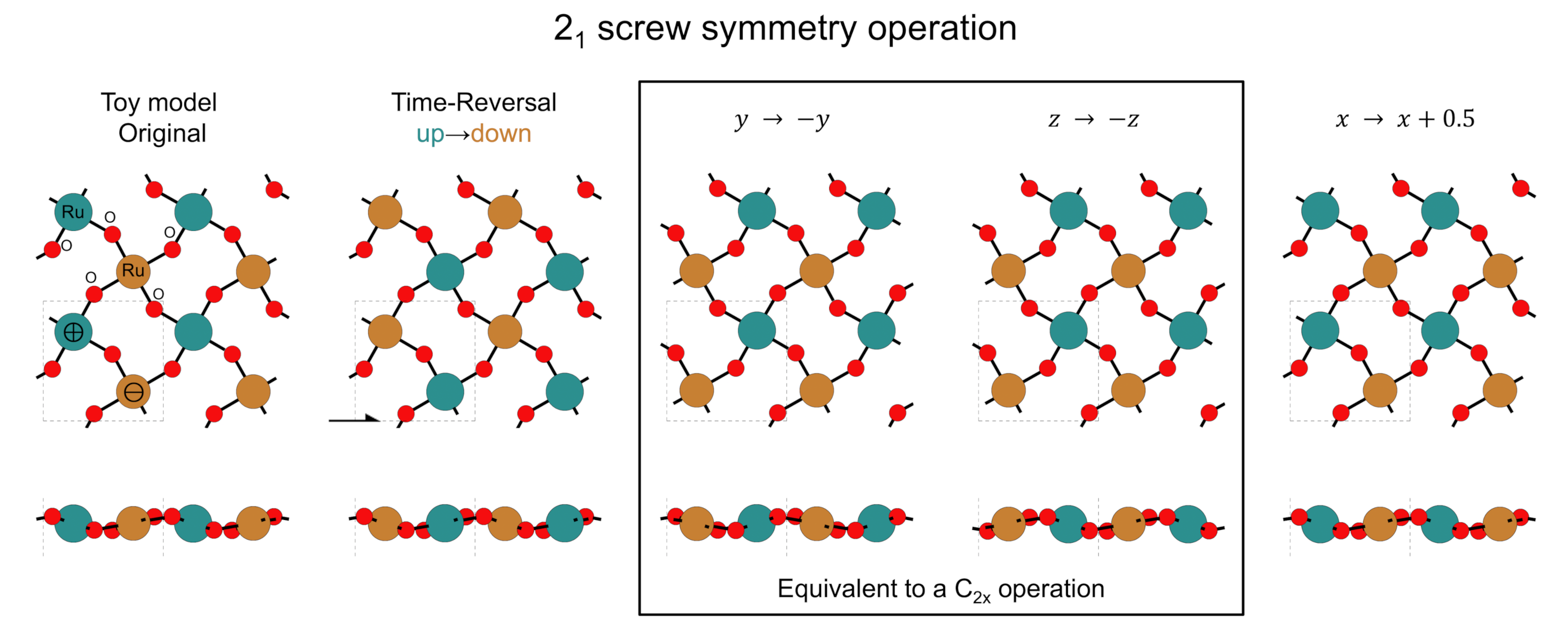} 
\caption{Symmetry operations responsible for the altermagnetic character of the toy model. Starting from the Ru-O plane of the 2D-CRO (first panel), a time-reversal operation is applied (second panel). Subsequently, a $C_{2x}$ operation is performed (third and fourth panels), followed by a translation in the x-direction (fifth panel).}
\label{Fig:toysim}
\end{figure}
 
As discussed in the main text, our toy model corresponds to the Ru-O plane of the 2D-CRO; we identify this core layer as the key component that generates the altermagnetic character of 2D-CRO. Figure \ref{Fig:toysim} presents the symmetry operations responsible for such altermagnetic character. Note that the same symmetry operations apply to 2D-CRO, although it is more difficult to follow the results of the different operations.

In the first panel, we present our starting point, with four Oxygen atoms (depicted as red spheres) around each Ruthenium atom (green for the positive magnetic moment and orange for the negative magnetic moment). The second panel results from a time-reversal operation, corresponding to an inversion of the magnetic moments.

The task now is to find a way to recover the initial configuration (first panel) by applying symmetry and translation operations on the structure affected by the time-reversal operation (second panel). 
After a symmetry analysis using the PYMATGEN library\cite{ong2013python}, we found that the $2_1$ screw symmetry operation achieves this. The $2_1$ screw operation can be separated into two stages: a rotation followed by a translation. The rotation corresponds to a $C_{2x}$ operation, equivalent to the $y \rightarrow -y$ inversion followed by the $z \rightarrow -z$ inversion, shown in the third and fourth panels. Finally, a translation of half a unit cell in the $x$-direction allows us to recover the initial configuration, as illustrated in the fifth panel.

\begin{figure}[h!]
\centering
\includegraphics[clip,width=.7\columnwidth,angle=0]{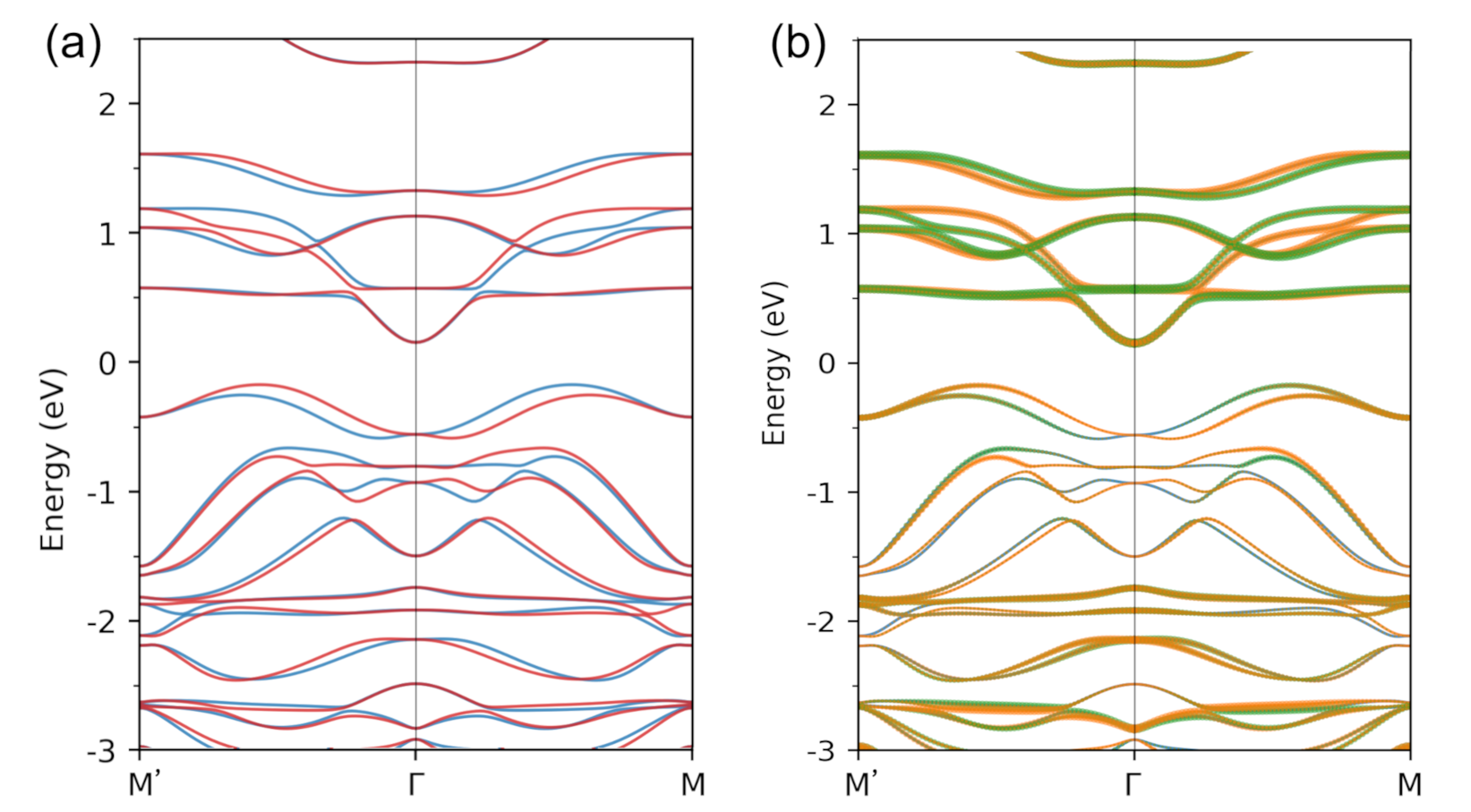} 
\caption{(Color online) The toy model band structure. In (a), the projection of spin-up and spin-down components is shown with blue and red lines, respectively; this panel is the same as Fig. \ref{Fig:toy_bands}. In (b), the projection onto the Ru atoms is depicted, where green and orange lines represent the projections onto the two different Ru atoms.
}
\label{Fig:toy_pband}
\end{figure}

\section{Phonon dispersion bands}
\begin{figure}[h!]
\centering
\includegraphics[clip,width=0.45\columnwidth,angle=0]{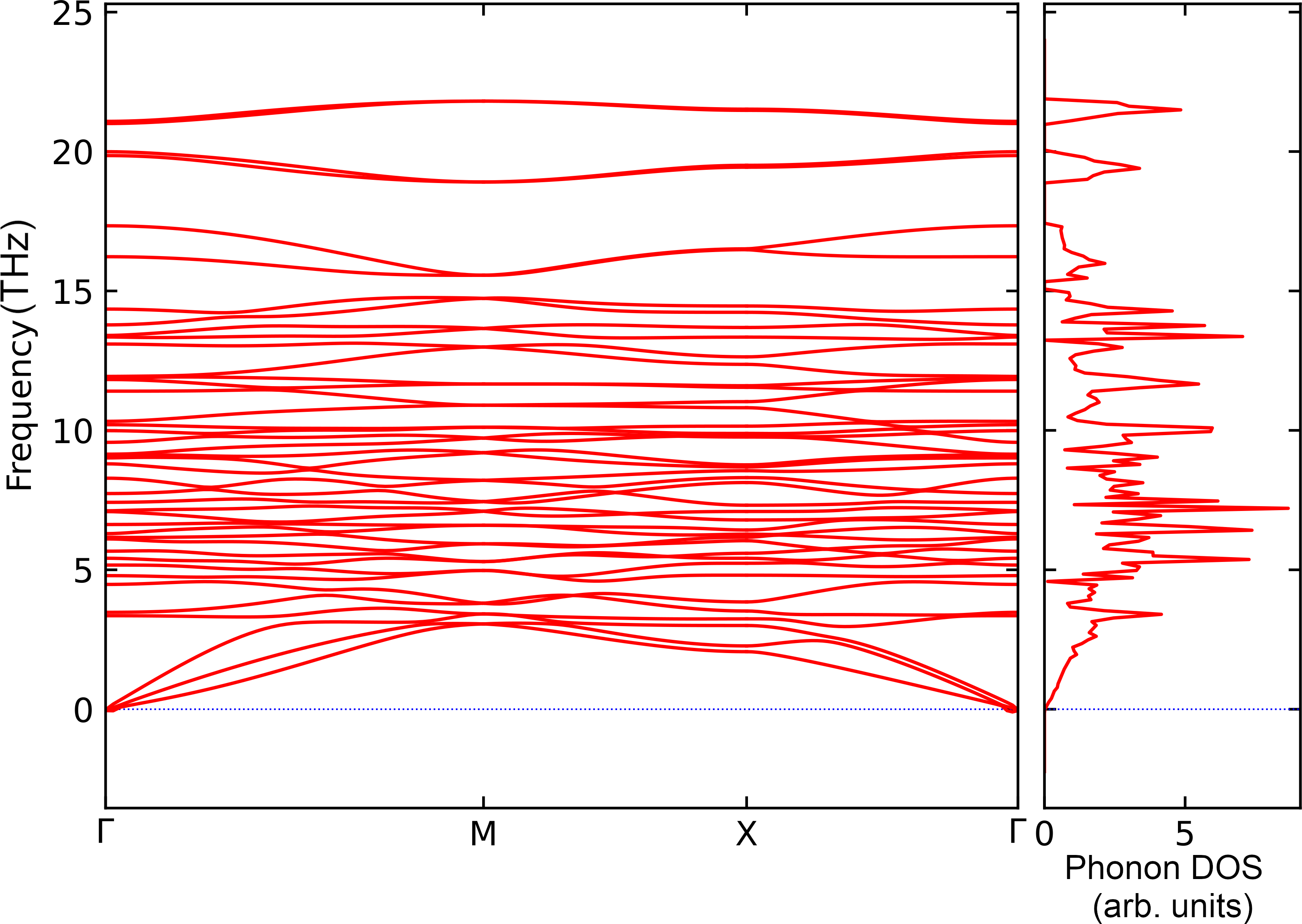} 
\caption{Phonon dispersion bands and density of states (DOS) of 2D-CRO for U$= 1$ eV. The absence of negative frequencies indicates the absence of imaginary phonon modes, confirming the mechanical stability of the system. 
}
\label{Fig:phonons}
\end{figure}

The phonon band structure (in Fig. \ref{Fig:phonons}) along the high-symmetry points shows no values at negative frequencies, indicating stability.

\newpage

\section{Molecular dynamics}
\begin{figure}[h!]
\centering
\includegraphics[clip,width=.44\columnwidth,angle=0]{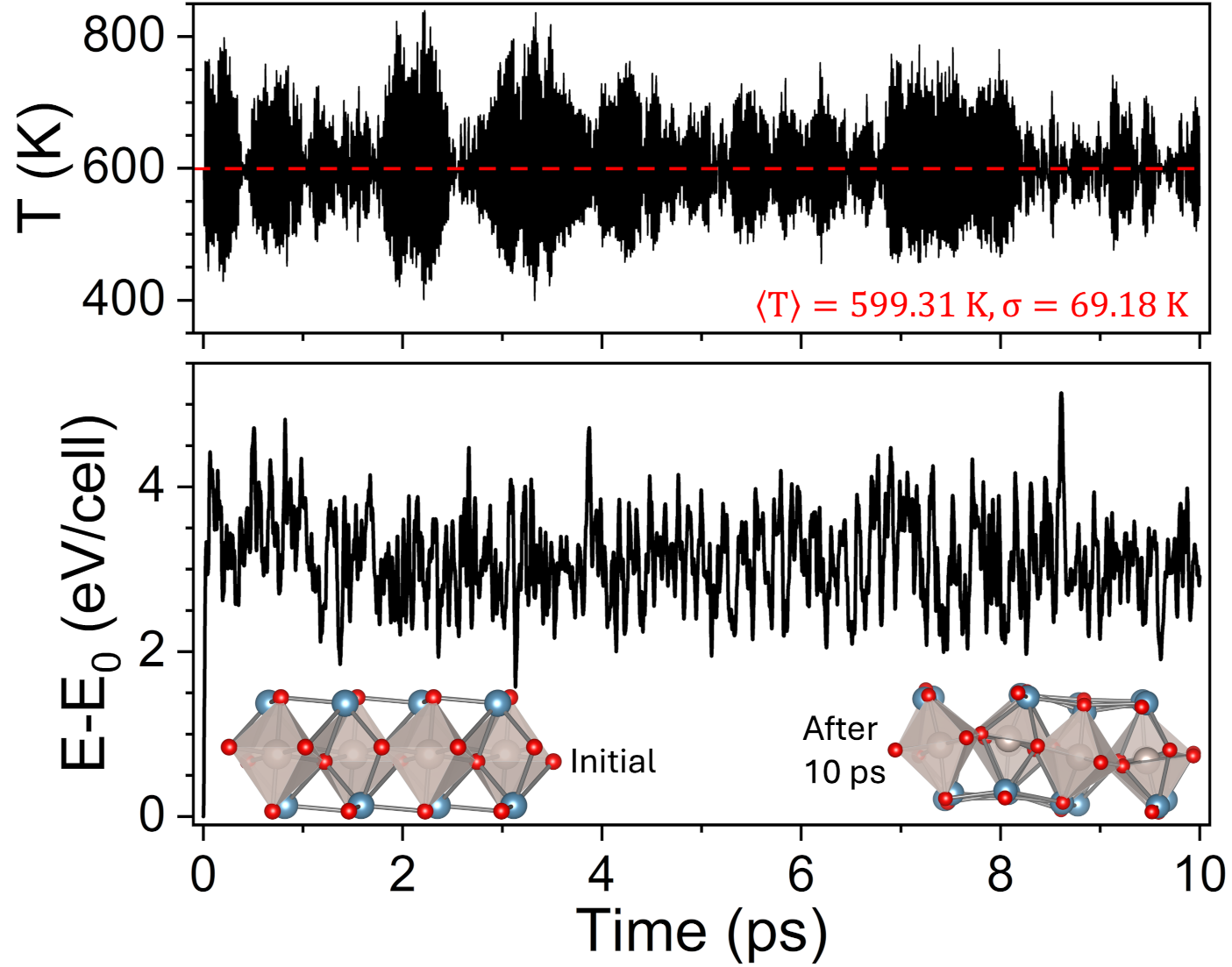} 
\caption{(Color online) Temperature and energy variation over time in molecular dynamics 
simulations of the $2\times2\times1$ cell using a Nos\'e-Hoover thermostat at 600 K. 
Panel (a) shows the system temperature, and panel (b) shows the energy variation 
relative to the initial configuration ($E_0$). The inset in panel (b) depicts the 
initial and final geometries after 10 ps at 600 K. The observed stability in both 
energy and geometry confirms the dynamical stability of the monolayer at this temperature. 
}
\label{Fig:MD}
\end{figure}

In Figure \ref{Fig:MD}, energy variation and temperature evolution for a $2\times2\times1$ supercell for the 2D-CRO in FM configuration with U = 1 eV. The system maintains structural stability after 10 ps at 600 K, only showing small bending due to thermal agitation.

\section{Elastic properties \label{sec:elastic}}
\begin{figure}[h!]
\centering
\includegraphics[clip,width=0.48\textwidth,angle=0]{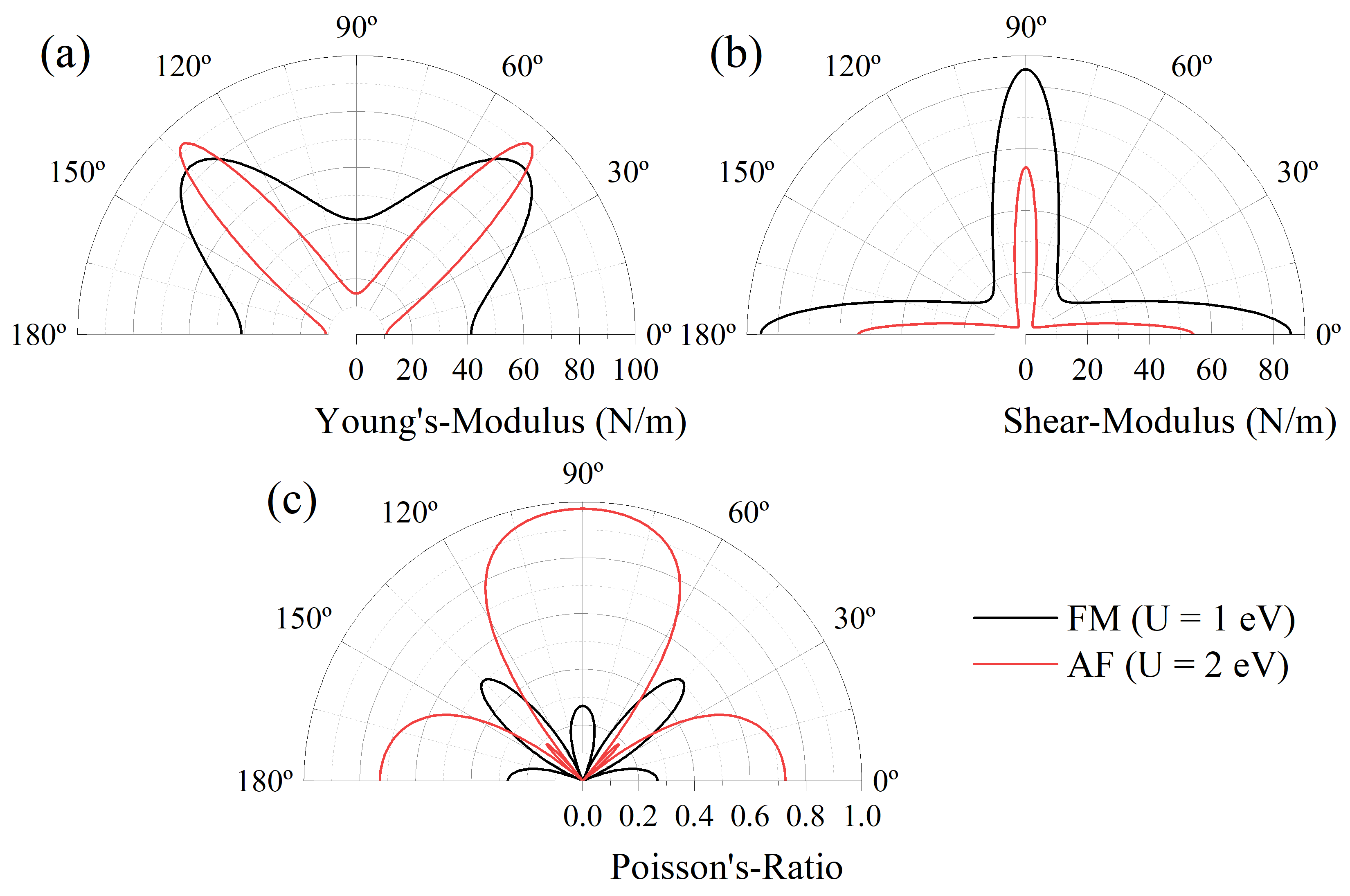} 
\caption{(Color online) Anisotropic mechanical properties of the 2D-CRO monolayer. (a) Young's modulus, (b) shear modulus, and (c) Poisson's ratio are plotted as a function of the angle in the ab-plane. The black lines represent the ferromagnetic (FM) configuration with U = 1 eV, while the red lines represent the antiferromagnetic (AF) configuration with U = 2 eV. Note the directional dependence of the mechanical properties and the impact of different magnetic configurations on the elastic properties.}
\label{Fig:elastic}
\end{figure}

Figure \ref{Fig:elastic} shows the main elastic properties of the 2D-CRO for the FM (U = 1 eV) and AF (U = 2 eV) configurations. 
The elastic properties exhibit angular dependence\cite{wang2022high}. For the ferromagnetic configuration (U = 1 eV), Young's modulus varies between 41.18 N/m and 84.67 N/m, the shear modulus ranges from 16.24 N/m to 85.56 N/m, and Poisson's Ratio fluctuates between -0.51 and 0.27.
In the antiferromagnetic configuration  (U = 2 eV), Young's modulus spans from 10.91 N/m to 92.39 N/m, the shear modulus changes between 3.41 N/m and 54.05 N/m, and Poisson's Ratio shifts from -0.2 to 0.98.

\end{document}